\documentclass[aps,prb,twocolumn,superscriptaddress]{revtex4-2}
\usepackage[T1]{fontenc}
\usepackage{amsmath}
\usepackage{amssymb}
\usepackage{graphicx}
\usepackage{hyperref}
\renewcommand{\vec}[1]{\mathbf{#1}}

\begin{document}

\title{Ground States and Excitations of 
Magnetic Impurities \\
in Pseudogap Superconducting Systems}

\author{Miguel A. Cazalilla}
\affiliation{Donostia International Physics Center (DIPC), 20018 Donostia-San Sebastian, Spain}
\affiliation{Ikerbasque, Basque Foundation for Science, 48013 Bilbao, Spain}
 
\author{Chen-How Huang}
\affiliation{Department of Physics and Nanoscience Center, University of Jyväskylä, P.O. Box 35 (YFL), FI-40014 University of Jyväskylä, Finland}

\begin{abstract}
 Combining effective field theory and  numerical renormalization-group (NRG),  we study the ground-state phase diagram and single-particle excitations of 
 a spin-$\tfrac{1}{2}$ impurity in a  superconducting system with a
tunneling density of states behaving as $\rho(\epsilon) \sim |\epsilon|^{r}$, 
 for $|\epsilon|\gg \Delta$ ($\Delta$ being the $s$-wave pairing potential). We focus on the properties of the doublet-singlet transition at large Kondo coupling. 
 The effective field theory for the singlet phase is inferred from a strong coupling
 expansion in the Kondo coupling. For $\Delta \neq 0$, it contains a local pairing
 term which drives the system into a spin-singlet phase with enhanced paring  correlations. We study how the singet-doublet phase boundary is affected by  particle-hole symmetry breaking perturbations such as a scattering potential and/or the chemical potential. Results for the $T$-matrix spectral function are also reported near the transition both at particle-hole symmetry and away from it. It is  shown that the singlet-doublet transition can be induced by the chemical potential rather than the Kondo coupling strength. At particle-hole symmetry, a resonance-like feature is observed for $r= 1$ and related to a two-quasiparticle excitation using a single-site model which is derived from  effective field theory. 
\end{abstract}
\date{\today}
\maketitle

\section{Introduction}

 Compared to their (state-of-the-art) semiconducting counterparts~\cite{DiVincenzo_PhysRevA.57.120,Burkard_RevModPhys.95.025003}, 
 superconducting spin qubits (SCSQs)~\cite{Nazarov_PhysRevB.81.144519,Nazarov2_PhysRevLett.90.226806,PitaVidal2023,pitavidal2025review}  exhibit rather short coherence times. However, besides allowing to access and manipulate the electron spin, the state of such superconducting systems is characterized by an additional quantum number, namely  the fermion parity, $P$.  By combining multiple SCSQs,  it is possible to design minimal Kitaev chains~\cite{Sau2012,Leijnse2012,Deng2016,Dvir2023,Bordin2025} that realize a
 decoherence-free quantum memory.  A standard paradigm for SCSQs requires  the many-body ground state of  a quantum dot coupled to an adjacent superconductor (SC), which together localize
 a spin-$\tfrac{1}{2}$  moment. In the absence of spin-orbit interactions~\cite{Zazunov_PhysRevLett.103.147004,Sau2012,PitaVidal2023} or applied magnetic fields~\cite{Leijnse2012,Sau2012}, the dot-SC system  forms an energy-degenerate spin doublet, or local moment (LM). 

However, a major obstacle for the stability of the LM is Kondo screening~\cite{Hewson_1993,Anderson_1970,Wilson_RevModPhys,Bulla_RevModPhys.80.395}. As the exchange coupling $J$ between the dot and the SC increases---e.g. by making dot-SC barriers more transparent deep in the Coulomb-blockade regime~\cite{Glazman_Pustilnik_2004}---the system inevitably crosses a critical threshold, $J_c$. Beyond $J_c$, the LM is erased and replaced by a spin-singlet ground state~\cite{Satori1992,SHIBA1993239,Bulla_RevModPhys.80.395}. This transition is not merely a change in magnetic response. It is a level-crossing transition that flips the ground-state fermion parity. In the vicinity of the transition, the singlet-doublet energy gap narrows drastically, severely exposing the system to quasiparticle poisoning due to low-energy $P$ fluctuations~\cite{Aumentadoetal2023,Pan2022Engineering}.

To circumvent these problems, it is crucial to find ways to push the critical coupling $J_c$ to higher values. One of the goals of this work,
is to expand our previous study on how  the host tunneling density of states (TDOS)---or an energy-dependent tunneling amplitude into the dot~\cite{Bulla_RevModPhys.80.395}---can suppress Kondo screening and enhance $J_c$. We focus specifically on  ``superconducting pseudogap systems'' (SCPS). These are host materials characterized by a power-law TDOS taking the form $\rho(\epsilon) \sim |\epsilon|^r$ with $r > 0$ above the $s$-wave pairing potential  energy scale, $\Delta$. 

  SCPS are not merely theoretical constructs; prominent examples for $r = 1$ include  graphene~\cite{WangJ_PhysRevB.98.121411,Trivini2025} and topological insulators  proximitized by conventional SCs~\cite{FuKane2008,Stanescuetal2010,Wangetal2012,Xuetal2014,Loss_PhysRevB_2014,Trivini2025}, as well as materials exhibiting $d+is$-wave pairing. The latter  has recently emerged as a candidate for the pairing symmetry in infinite-layer nickelates~\cite{Gu2020,Ji_NatComm2023,Normand_PhysRevB.110.024514,Wang_PhysRevB.102.220501}. See Appendices~\ref{app:scps} and \ref{app:scps2} for details of the derivation of the local Green's function for both types of systems.

 A shown in  Appendix~\ref{sec:ysr},   the ``classical'' approach of Yu~\cite{Yu}, Shiba~\cite{Shiba}, and Rusinov~\cite{Rusinov} (YSR), when extended to  SCPS,  qualitatively hints at an enhanced $J_c$ compared to metallic hosts. For instance, for $r=1$ and $\Delta/D = 10^{-3}$, it yields $J_c/D \sim 10$, compared to $J_c/D \sim 1$ for $r=0$~\cite{Loss_PhysRevB_2014}. However, because the YSR approach neglects both the quantum fluctuations of the impurity spin and the many-body physics of Kondo screening, it cannot be reliably  trusted.   In the companion letter~\cite{unpub}, we have explored the ground state phase diagram both analytically and  using the numerical renormalization group (NRG). 
 Thus, we have  shown that,  while the low-energy depletion of single-particle states in SCPS weakens Kondo screening, the spin of the ground state at strong coupling is determined by a complex competition between Andreev reflection  and the Kondo-singlet polarization energy. Consequently, even at particle-hole symmetry (PHS), the physics governing the ground state of a spin-$\tfrac{1}{2}$ impurity in a SCPS differs from the standard pseudogap Kondo problem~\cite{Fradkin_PhysRevLett.64.1835,Chen_1995,Ingersent_PhysRevB.57.14254,Ingersent_PhysRevB.57.14254,Fritz_Vojta_2004}.

 In this work,  we expand  on the results reported in the companion letter~\cite{unpub}, which mainly dealt with the ground-state phase diagram for systems at or near PHS. We also provide the full details of the derivation of the effective field theory  describing the impurity system at strong coupling for both normal (cf Sec.~\ref{sec:normal}) and superconducting systems (cf. Sec.~\ref{sec:super}). In addition,  in Sec.~\ref{sec:super}  we  introduce a renormalized  single-site model in order to describe the competition between the Andreev reflection and the Kondo-singlet polarization energy that determines the ground state for $r > 1/2$ (see Sec.~\ref{sec:super}). In Sec.~\ref{sec:nrg}, we present additional NRG results  for the  phase diagram (cf. subsection~\ref{subsec:phasediag}) over a wide range of values for the scattering potential and the chemical potential. Finally, in Sec.~\ref{subsec:specfun}, we discuss the NRG results for the $T$-matrix spectral function both at PHS and at finite chemical potential. In the latter case, we illustrate how  in SCPS  the doublet-singlet transition can be induced by the changing the chemical potential. Finally, the Appendices contain the derivation of number of results that are employed in various analytical derivations in the article, as well as the technical details of  the NRG calculations.

\section{Model and definitions}
\begin{figure}[b]
    \centering
    \includegraphics[width=\columnwidth]{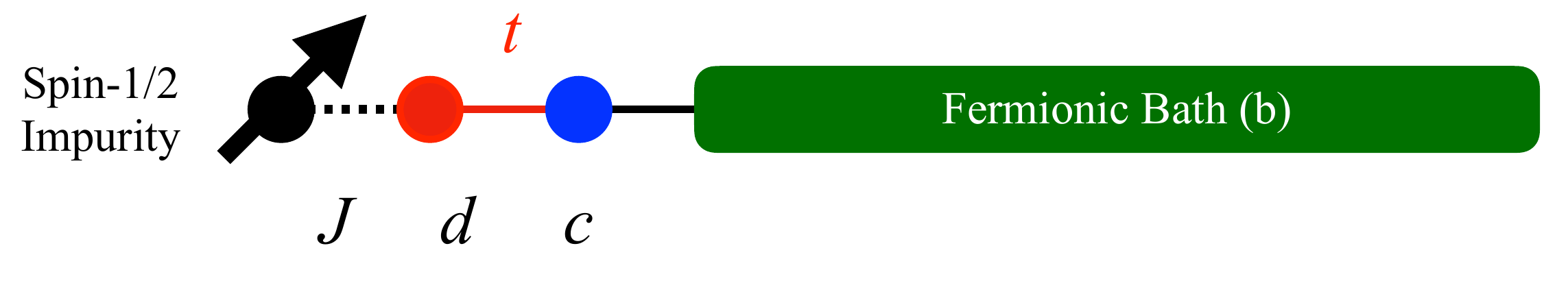}
    \caption{(a): Sketch of the modified Kondo model(s) studied in this work. It consists of a quantum spin-$\tfrac{1}{2}$ impurity coupled to site ($d$) via anti-ferromagnetic (Kondo) exchange $J>0$. The $d$-site is coupled to the $c$ site via a hoping amplitude $t$. Finally, the $c$-site is coupled to a fermionic bath.}
    \label{fig:fig1_sketch}
\end{figure}
 We shall consider   a  spin-$\tfrac{1}{2}$ quantum impurity
 (e.g. a quantum dot in the Coulomb blockade regime with an odd number of electrons and spin $S=\tfrac{1}{2}$ in the ground state~\cite{Glazman_Pustilnik_2004}) coupled to a system of itinerant fermions (referred to as the `host') via a Kondo exchange coupling $J > 0$. The host 
may be a normal or superconducting and it is characterized by a tunneling density of states (TDOS) $\rho(\epsilon)$ that displays a pseudogap behavior, i.e. $\rho(\epsilon) \sim |\epsilon|^{r}$ for $|\epsilon|\gg \Delta$, where $\Delta$ is the $s$-wave pairing potential ($\Delta = 0$ in the ``normal'' case).  We further assume that only one scattering channel  couples to the impurity. After elimination of the impurity (quantum dot's) levels~\cite{Glazman_Pustilnik_2004}, one arrives at the following Kondo model (cf. Fig.~\ref{fig:fig1_sketch}):
\begin{align}
H &= H_I +H_T +  H_{c} +  H_b-\mu N,\notag\\
H_I &= J  \vec{S}\cdot \left( d^{\dag}_{\sigma}   \vec{s}_{\sigma,\sigma^{\prime}} d_{\sigma^{\prime}} \right) \notag
\\ &\qquad + V d^{\dag}_{\sigma} d_{\sigma} + \Delta \left( d^{\dag}_{\uparrow} d^{\dag}_{\downarrow} + \mathrm{H.c.}\right),\notag\\
H_T &= -t \left( d^{\dag}_{\sigma} c_{\sigma} +  \mathrm{H.c.} \right),\notag\\
H_{c}&=   \Delta \left( c^{\uparrow}_{\uparrow} c^{\dag}_{\downarrow} + \mathrm{H.c.} \right),
\label{eq:mKm}
\end{align}
where $N = d^{\dag}_{\sigma}d_{\sigma}+ c^{\dag}_{\sigma}c_{\sigma} + N_b$ is total number operator. 
In  the previous expressions and the ones that follow, we shall use Einstein's convention which implies  summation over repeated spin indices i$\sigma = \uparrow,\downarrow$ of the Fermi operators for the $d,c$-fermions, i.e. $d_{\sigma},c_{\sigma}$. Moreover, $\vec{s} = \frac{1}{2} (\sigma^x, \sigma^y, \sigma^z)$, where $\sigma^{x,y,z}$ are the spin Pauli matrices and $\vec{S} = (S^x,S^y,S^z)$ are the spin-$\tfrac{1}{2}$ operators of the impurity.
The Hamiltonian $H_b$ describes a \emph{non-interacting} fermionic `bath' of bandwidth $D \simeq t$. If necessary (as it is the case for calculations using the numerical renormalization group, NRG), it can be represented as a tight-binding Hamiltonian of the form:
\begin{align}
H_{b}- \mu N_b &=  \sum_{m=1}^{L-1} \left[ -t_m \left( b^{\dag}_{m,\sigma} b_{m+1,\sigma} + \mathrm{H.c.} \right) 
+ \xi_m b^{\dag}_{m,\sigma}b_{m\sigma}\right] \notag\\
&\, + \Delta \sum_{m} \left(b^{\dag}_{m,\uparrow} b^{\dag}_{m,\downarrow} + \mathrm{H.c.} \right), \label{eq:tb}
\end{align}
where $b_{m\sigma}$ ($b^{\dag}_{m\sigma}$) are fermion destruction operators at site $m=1,\ldots, L-1$ and spin projection $\sigma$. The bath and $c$-fermion site Hamiltonians depend on several parameters such as the $s$-wave pairing potential $\Delta$ and the chemical potential $\mu$, 
hoping amplitudes $\{t_m\}_{m=1}^{L-1}$ and $t$, and the onsite energies $\{\xi_m = \epsilon_m - \mu\}_{m=1}^{L-1}$, which 
encode properties such as the pseudogap TDOS. 
Nevertheless, rather that working with $H_b$ in the form of Eq.~\eqref{eq:tb}, for the analytical calculations that follow we shall often describe the `bath' in terms of 
the  (Fourier transform of the) local Matsubara Green's function of a non-interacting chain ending at the  $c$-fermion site. The latter describes the effect of the bath degrees of freedom by means of the following (Nambu matrix) correlation function:
\begin{equation}
\mathcal{G}_{0c}( i\omega_n) = - \int^{\beta}_0 d\tau \, e^{i\omega_n \tau} \langle \mathcal{T} \left[ \Psi_c(\tau) \otimes 
\Psi^{\dag} (0)_c \right] \rangle_B,\label{eq:greenc}
\end{equation}
where $\Psi^T_c = \left(c_{\uparrow}, c^{\dag}_{\downarrow}\right)$ is a Nambu spinor describing the electrons at the $c$-fermion site, $\otimes$ is the dyadic product, 
 $\beta = 1/T$ is the inverse absolute temperature, and  $\omega_n = (2n+1) \pi/\beta$ are the fermionic Matsubara frequencies.
In Eq.~\eqref{eq:greenc}, $\langle\ldots \rangle_B$ means that only the bath degrees of freedom are traced out.  Setting $t = 0$ and using the recursion method to integrate all bath sites, $\mathcal{G}^{0}_c( i\omega_n)$ can be written as ``continued faction'':
\begin{multline}
    \mathcal{G}_{0c}(i\omega_n) = \left[ \omega_n - \mu \tau^3 - \Delta \tau^1 -  t^2_1 \tau^3  \left[ i\omega_n - \xi_1 \tau^3 - \Delta \tau^1 \right. \right. \\ 
  \left.  -  t^2_2 \tau^3  \left[i\omega_n - \xi_2 \tau^3  
-\Delta \tau^1  -  \cdots\right]^{-1} \tau^3   ]^{-1}   \tau^3 \right]^{-1},
\end{multline}
where $\tau^{1,2,3}$ are the $2\times 2$ Pauli matrices acting on the indices of the Nambu spinors.

An alternative representation of the host (including the $d$-, $c$-, and  the bath  sites) using an effective \emph{chiral} one-dimensional Fermi system with power-law dispersion is introduced in Appendix~\ref{app:cmodel}. Nevertheless, since we are interested in the properties of the system in the low-energy regime, for $\Delta,|\omega_n|\ll D$, the microscopic details of  bath representation are not  important and we shall often switch between different representations, depending on convenience.   

In the strong coupling limit to be discussed below where  $J\gg t\simeq D$,  the $d$-fermion becomes strongly entangled with the impurity in a Kondo singlet (KS) state,  which described by the following ket: 
\begin{equation}
|\Phi_0\rangle =  \frac{1}{\sqrt{2}} \left[ |\Uparrow\rangle \otimes \gamma^{\dag}_{\downarrow}|BCS\rangle - |\Downarrow\rangle   
\otimes \gamma^{\dag}_{\uparrow}|BCS\rangle \right], \label{eq:ks}
\end{equation}
where the $\gamma_{\sigma}$ are the Bogoliugov quasi-particle operators related to $d_{\sigma}$ and $d^{\dag}_{\sigma}$ a Bogoliubov transformation (see Eq.~\eqref{eq:gammabogol} in Appendix~\ref{sec:zeroBW}). The state $|BCS\rangle$ is also defined in Appendix~\ref{sec:zeroBW}.
Note that 
$|KS\rangle$ is the ground state of the 
Hamiltonian $H_I$ (cf. Eq.~\ref{eq:mKm}) describing 
the magnetic impurity coupled to the $d$-fermion site for $J > 4 \sqrt{\Delta^2+V^2}/3$ (at $\mu  = 0$). For smaller values of $J$, the ground state is a spin doublet where the $d$-fermions remain paired in a BCS state. 

 In the strong coupling KS regime,  the magnetic  impurity becomes `screened'.  The $d$-fermions are also `eliminated' as no quasi-particle can effectively hop onto the $d$-site. The resulting system is apparently non-magnetic. 
 However, as Wilson~\cite{Wilson_RevModPhys} and Nozi\`eres~\cite{Nozieres_1974}  pointed out,
 the system that is left behind by the formation of the KS does not  behave like an ordinary, non-magnetic
 impurity and it is indeed a  interacting system, which can be described by an effective field theory whose derivation is provided below. 

\section{Normal Fermi Systems}\label{sec:normal}

In this section, in order to establish the connection with Nozieres' \emph{local} Fermi-liquid
theory~\cite{Nozieres_1974},  we shall first consider the case of a normal Fermi system  (with pseudogap TDOS). 
Thus, we set $\Delta = 0$ in  Eqs.~\eqref{eq:mKm} and \eqref{eq:tb}.  Another reason to consider the normal
Fermi system case first is that it is gapless and therefore allows to rigorously  define renormalization-group (RG) transformations as well as the scaling dimension of the various couplings appearing in the
effective action to be discussed below. In addition to setting $\Delta = 0$, for the sake of simplicity we also assume  that $\mu=0$ and $V$ is not too large so as to substantially modify the TDOS (thus we set $\xi_m  = 0$ for $m = 1, \ldots, L$ in the tight-binding representation of $H_b$, Eq.~\eqref{eq:tb}). For the model defined by Eqs. \eqref{eq:mKm} and  \eqref{eq:tb}, we recall that PHS  is defined as invariance under the following transformation :
 \begin{align}
 d_{\sigma} &\to  \,\, d^{\dag}_{-\sigma}, \\
 c_{\sigma} &\to -c^{\dag}_{-\sigma}, \\
 b_{m,\sigma} &\to (-1)^{m+1} b^{\dag}_{m,-\sigma},  
 \end{align}
which leaves all terms in the Hamiltonian unchanged except the potential scattering $\propto V$ which changes sign $V \to -V$ (along with chemical potential $\mu\to -\mu$).

A key assumption to be made below is that the effective field theory for Eq.\eqref{eq:mKm} to be derived below in the strong coupling regime where $J \gg D\simeq t$ remains valid at smaller values of $J\lesssim D$. This is certainly the case for Nozi\`eres' local Fermi liquid theory~\cite{Nozieres_1974} for constant TDOS  (i.e. $r = 0$), where it applies for any $J > 0$. Whether this remains true for arbitrary values of $r$ and $J$ needs to be assessed \emph{a posteriori} using the renormalization group, and  depends on the additional symmetries of each particular realization of \eqref{eq:mKm} such as PHS or, as we shall see in the following section, on the existence of a $s$-wave pair potential $\propto \Delta$.

  In order to obtain the low-energy description of the KS, we generalize the strong coupling approach invoked by Nozi\`eres to systems with an arbitrary pseudo-gap tunneling
 density of states (TDOS). In the following section, we shall see that the same approach can also be applied
 to systems with a pairing gap in the single-particle spectrum.  Although it is possible to formulate the strong coupling approach in the operator 
formalism, since the approach described below requires  integrating out the degrees of freedom of the  bath together with the $d$-fermion
site that forms the KS with the impurity, the path integral formalism is more convenient.  Thus, the partition function is  written as an functional integral over a Grassmannian algebra and the impurity spin:
\begin{equation}
Z = Z_b \int \mathcal{D}[\vec{S} \bar{d} \bar{c}  d  c]  \, e^{S\left[\vec{S},\bar{d},\bar{c},d,c\right]}\label{eq:part}
\end{equation}
where the prefactor $Z_b$ in ~\eqref{eq:part} is the partition function of the bath. The action $S\left[\vec{S},\bar{d},\bar{c},\bar{b},d,c,b\right]$ reads:
\begin{align}
S &= S_{i}[\vec{S}] + S_{I}\left[\vec{S},\bar{d},d\right] + S_{dc}\left[\bar{c},c\right] + S_{T}\left[\bar{c},\bar{d},c,d\right],
\notag\\ 
S_{dc}  &= -\frac{1}{\beta}\sum_{\omega_n} \left[ \bar{d}_{\sigma} \left(-i\omega_n - \mu \right) d_{\sigma}  +   \bar{c}_{\sigma}  G^{-1}_{0c} ( i\omega_n ) c_{\sigma} \right] ,\label{eq:sc}\notag\\
S_{I}  &= S_K + S_V =    \int d\tau\,  \left[ J \vec{S} \cdot \left( \bar{d}_{\sigma} \vec{s}_{\sigma,\sigma^{\prime}}  d_{\sigma^{\prime}} \right) + V \bar{d}_{\sigma} d_{\sigma} \right] ,\notag\\
S_{T} &= -\frac{t}{\beta}\sum_{\omega_n} \left( \bar{d}_{\sigma} c_{\sigma} + \bar{c}_{\sigma} d_{\sigma} \right). 
\end{align}
Since the bath is non-interacting, i.e. its action is quadratic, it can be integrated out exactly leading to Eq.~\eqref{eq:sc}, where $G_{0c}(\omega_n)$ is the only independent diagonal component of the Nambu matrix~\eqref{eq:greenc} for $\Delta=0$, i.e.
\begin{equation}
G_{0c}( i\omega_n) = - \int^{\beta}_0 d\tau \, e^{i\omega_n \tau} \langle \mathcal{T} \left[ c(\tau)  c^{\dag} (0) \right] \rangle_B.\label{eq:greenc2}
\end{equation}
 Regarding the action for the impurity spin, in the case of spin-$\tfrac{1}{2}$ we can rely on, e.g. the Popov-Fedotov fermionic representation~\cite{Popov_Fedotov_1988}, for which 
$\vec{S} = \bar{f}_{\sigma} \vec{s}_{\sigma,\sigma^{\prime}} f_{\sigma^{\prime}}$, where
the Grassmann fields $\bar{f} = \bar{f}( \omega_n)$ and $f = f( \omega_n)$ have the following action, which yields the following
series for the resulting effective action:
\begin{equation}
S_{f} = S_f\left[ \bar{f},f\right] = \frac{1}{\beta} \sum_{\omega_n} \bar{f}_{\sigma} \left(-i \omega_n  - \mu_f \right) f_{\sigma},\label{eq:spinferm}
\end{equation}
where $\mu_f = -i \pi/(2\beta)$ and we have used Einstein's convention for the summation over repeated (spin) indices $\sigma,\sigma^{\prime}$. Thus, the Kondo coupling takes the form:
\begin{equation}
S_{K} = J \int^{\beta}_0 d\tau  \left( \bar{f}_{\sigma} \vec{s}_{\sigma,\sigma^{\prime}} f_{\sigma^{\prime}} \right) 
\cdot \left( \bar{d}_{\lambda}  \vec{s}_{\lambda,\lambda^{\prime}}  d_{\lambda^{\prime}} \right),\label{eq:sKf}
\end{equation}
 Next, we shall integrate out  the KS formed by the magnetic impurity and the $d$-fermion. Unlike the integration of the bath,  this cannot be done
exactly because the Kondo exchange   effectively describes an interaction  between the $d$- and $f$-fermions (cf. Eq.~\ref{eq:sKf}). Nevertheless,  in the large $J\gg t\simeq D$ regime it is possible to make analytical progress by using the cumulant expansion:
\begin{align}
S_{KS}^{\mathrm{eff}} &= -\log \: \langle e^{-S_T} \rangle_{KS} \\
&= -\frac{1}{2!} \langle S^2_T \rangle^c_{KS}  - \frac{1}{4!} \langle S^4_T\rangle^c_{KS}    + \ldots
\end{align}
where $\langle S^2_T \rangle^c_{KS} = \langle S^2_T \rangle_{KS}$ and $ \langle S^4_T\rangle^c_{KS} =  
\langle S^4_T\rangle_{S} -  3 \langle S^2_T\rangle^2_{KS} $, where $\langle \ldots \rangle_{KS}$ stands for the average over the KS ground-state expectation  of the (products of the) $d$ operators (see Appendix~\ref{sec:zeroBW} for details).  The leading-order terms in the above expansion are $O(t^2)$ and $O(t^4)$
and  yield the following effective action for the $c$-fermion site:
\begin{align}
S &= S_0 + S_1,\\
S_0 &=  \frac{1}{\beta} \sum_{\omega_n} \bar{c}_{\sigma} \left[ -G^{-1}_{0c}(i \omega_n) 
+ \Sigma_c(i \omega_n) \right]  c_{\sigma},\label{eq:s01}\\
S_1  &= -\frac{1}{4} \int  d\tau_{1}d\tau_{2}d\tau_{3}d\tau_{4}\:   \bar{c}_{\sigma_1}(\tau_1) \bar{c}_{\sigma_2}(\tau_2) \notag\\
&\quad \times \Gamma^{\sigma_1\sigma_2,\sigma_3\sigma_4}(\tau_1,\dots,\tau_4) c_{\sigma_3}(\tau_3) c_{\sigma_4}(\tau_2),
\end{align}
where 
\begin{align}
   \Sigma_c(i \omega_n) &= -t^2  \int^{\beta}_0 d\tau\, e^{i\omega_n \tau}\, \langle \mathcal{T} \left[ d_{\sigma}(\tau) d^{\dag}_{\sigma}(0) \right] \rangle_{KS},\notag\\
   &= t^2 G_{0}(i\omega_n),\\
 \Gamma^{\sigma_1\sigma_2\sigma_3\sigma_4}&(\tau_1,\dots,\tau_4)  = (-1)^2 t^4 \notag\\
 & \times \langle \mathcal{T} \left[  d_{\sigma_1}(\tau_1)  d_{\sigma_2}(\tau_2) d^{\dag}_{\sigma_3}(\tau_3) d^{\dag}_{\sigma_4}(\tau_4)\right] \rangle^c_{KS}. 
\end{align}
Thus, we obtain a self-energy and a time-dependent effective interaction on the $c$-fermion site.  The physical meaning of these terms is the  following:
 $\Sigma_c(i \omega_n)$ describes virtual  hopping processes in which the state of the $d$-site + impurity system
 fluctuates between a KS  and the spin-doublet states in which the $d$-site is either
 empty or doubly occupied (i.e. \emph{singlet-doublet} transitions). The vertex function $\Gamma^{\sigma_1\sigma_2,\sigma_3\sigma_4}(\tau_1,\dots,\tau_4)$
 describes an  interaction generated at the $c$-site via virtual hopping. In principle, it depends on imaginary times 
 in a complicated way. However, at low energies (i.e. for $|\omega_n| \ll J$), we expect it to reduce to a contact
 interaction of the form $\sim U_0 \bar{c}_{\uparrow} \bar{c}_{\downarrow} c_{\downarrow} c_{\uparrow}$. Thus,
 we can approximate ($\sigma =\uparrow,\downarrow$)
 \begin{multline}
  \Gamma^{\sigma,-\sigma,-\sigma,\sigma}(\tau_1,\tau_2,\tau_3,\tau_4) \simeq   -\frac{t^4}{J^2}
\delta(\tau_1-\tau_4)  \delta(\tau_2-\tau_3)  \\
\times C_{\uparrow,\downarrow}(\tau_1 -\tau_2),\label{eq:gamma}
 \end{multline}
 where
 \begin{equation}
  C_{\uparrow\downarrow}(\tau)= - \langle\mathcal{T} \left[  \left( n_{d,\uparrow}(\tau) - \tfrac{1}{2} \right) \left(  n_{d,\downarrow}(0) -\tfrac{1}{2}\right)\right] \rangle_{K}. \label{eq:cnn}
 \end{equation}
 Eq.~\eqref{eq:gamma} requires some additional explanation: The basic assumption is that the quantum fluctuations from the Kondo singlet to the triplet and back to the   singlet take place through intermediate (spin-doublet) states with energy $\sim J$. In the low energy limit, the latter    are essentially instantaneous (hence the Dirac delta in imaginary time) 
 and contribute a factor of $(-1/J)$ arising from the denominator in perturbation theory. The net effect is to yield a Hubbard-like
 interaction at the $c$-fermion site, which is mediated by the opposite-spin occupation fluctuations of the $d$-fermion site.  The latter are described by the correlation function 
 $C_{\uparrow\downarrow}(\tau)$ in the limit where  $J \tau\gg 1$. Note that  $C_{\uparrow \downarrow}(\tau)$   describes the
 fluctuations in the particle-hole channel only. We could  also consider fluctuations in the particle-particle channel, which are described by the correlation function $C_{\Delta}(\tau) = -\langle \mathcal{T} \left[ \Delta(\tau) \Delta^{\dag}(0)\right] \rangle$ where $\Delta = d_{\downarrow} d_{\uparrow}$. However, explicit calculation of this correlation function in the KS ground state of a $d$-site fermion and the impurity shows that $C_{\Delta} =0$. The reason is that $\Delta = d_{\downarrow} d_{\uparrow}$ is a 
 spin-singlet operator which does not couple the KS to the  spin-triplet states of the composite consisting of the $d$-fermion and the  impurity. In  Appendix~\ref{sec:zeroBW} we provide the details of the computation of the correlation functions $\Sigma(i\omega_n)$, $C_{\uparrow\downarrow}(i\omega_n)$ in the low frequency limit, i.e. for $|\omega_n|\ll J$. The resulting expressions are given 
 below, in Eqs.~\eqref{eq:selfnrg} and \eqref{eq:ccorr}.
 

In order to make further progress and  relate the action to the TDOS of the host at $J = 0$, we  rewrite the Green's function $G^{-1}_{0c}(i \omega_n)$ in~\eqref{eq:s01} in terms of  $G_{0d}(i \omega_n)$, which is  the Green's function at the $d$-fermion site of the chain (cf. Fig.~\ref{fig:fig1_sketch}) system for $J = 0$. Using the recursion relations,  
\begin{equation}
G_{0d}(i\omega_n) = \frac{1}{i\omega_n -  t^2 G_{0c}(i\omega_n)}.\label{eq:recur}
\end{equation}
Solving this equation for $G^{-1}_{0c}(i\omega_n)$ and substituting the result into Eq.~\eqref{eq:s01} yields
\begin{align}
S_0&=  
 \frac{1}{\beta} \sum_{\omega_n} \bar{c}_{\sigma} \left[ \frac{ t^2  G_{0d}(i \omega_n)}{1- i\omega_n G_{0d}(i \omega_n)} 
+ \Sigma(i \omega_n) \right]  c_{\sigma},\label{eq:seff}
\end{align}
where (see Appendix~\ref{sec:zeroBW} for details):
\begin{equation}
\Sigma(i \omega_n\ll J) \simeq - \left(\frac{4t}{3 J}\right)^2 (i\omega_n + V).  \label{eq:selfnrg}
\end{equation}
 Furthermore, using the approximation from Eq.~\eqref{eq:gamma}, we arrive at
\begin{multline}
S_1 \simeq \frac{t^4}{2 J^2} \int d\tau_1 d\tau_2 \: C_{\uparrow,\downarrow}(\tau_1-\tau_2) \\
\times \bar{c}_{\uparrow}(\tau_1) \bar{c}_{\downarrow}(\tau_2) c_{\downarrow}(\tau_2) c_{\uparrow}(\tau_1)\\
\simeq  U_0 \int d\tau\,   \bar{c}_{\uparrow}(\tau) \bar{c}_{\downarrow}(\tau) c_{\downarrow}(\tau) c_{\uparrow}(\tau).
\end{multline}
Eq.~\eqref{eq:gamma} is accurate as long as the integral over $\tau_1$ and $\tau_2$ is taken $|\tau_1-\tau_2|\gg J^{-1}$. 
In the second expression $U_0 = t^4/4J^3$, and we have used (see Appendix~\ref{sec:zeroBW} for details):
\begin{equation}
 C_{\uparrow\downarrow}(\tau) \simeq \frac{\delta(\tau)}{2J} \label{eq:ccorr}
\end{equation}
for $|\tau| \gg J^{-1}$.
Thus, to leading order at low energies the singlet-triplet fluctuations give rise to an effective Hubbard like interaction 
between the electrons residing on the $c$-site~\cite{Wilson_RevModPhys,Nozieres_1974,Chen_1995,Ingersent_PhysRevB.57.14254,Ingersent1_PhysRevB.54.11936,Fritz_Vojta_2004}.

The quadratic part of the effective action, Eq.~\eqref{eq:seff}, can be further simplified if we consider the low energy limit where $|\omega_n| \ll D\ll J$,  $D$ being the bandwidth. 
In this limit, $G_{0d}(i \omega_n) \sim -i |\omega_n/D|^r \mathrm{sgn}(\omega_n)$ (see Appendix~\ref{app:relation} for details). Hence, the low-energy effective action takes the form:
\begin{align}
S &= S_0 + S_1, \label{eq:sK}\\ 
S_0  &=-\frac{1}{\beta} \sum_{\omega_n} \bar{c}_{\sigma}  G^{-1}_c(i \omega_n)  c_{\sigma},\label{eq:s0d}\\
S_1 &= U_0 \int d\tau \left(\bar{c}_{\uparrow} c_{\uparrow} - \tfrac{1}{2} \right) \left( \bar{c}_{\downarrow} c_{\downarrow} - \tfrac{1}{2} \right),\label{eq:hubbard}
\end{align}
where  
\begin{equation}
G^{-1}_{c}(i \omega_n) = i A_0 \mathrm{sgn}( \omega_n) \left|\frac{\omega_n}{D}\right|^r + A_1 i\omega_n + V_1.
\label{eq:ginvc}
\end{equation}
In Eq.~\eqref{eq:s0d} and \eqref{eq:hubbard} we have written the interaction term $\propto U_0$ in a form that explicitly respects particle-hole symmetry. In the strong coupling regime where  
$J\gg D$,  the coefficients of the quadratic action are $A_0 = \pi \rho_0 t^2/\cos(\pi r/2)$,  $A_{1} =  (4t/3 J)^2$ for $r\neq 0$.
and $A_1 = A_0 +  (4t/3J)^2$ for $r = 0$, $V_1 = (4 t/3 J)^2 V$.  However, it is important to emphasize that the actual values of $A_0, A_1$ and $U_0$
can be very different in the most realistic situation where $J \lesssim D$, as already pointed out by Nozi\`eres~\cite{Nozieres_1974} for constant TDOS (i.e. $r = 0$). The effective theory has a built-in cutoff $\Lambda_K$, such that all Matsubara frequency summations are implicitly understood to be restricted by $|\omega_n| < \Lambda_K$. The cutoff $\Lambda_K < D$. For instance, for $r=0$, $\Lambda_K$ is proportional is of the order of the Kondo temperature $T_K$~\cite{Hewson_1993,Nozieres_1974}.

 The quadratic term of the effective $c$-fermion  action derived above corresponds to the saddle point
of a large-$N$ mean-field approach~\cite{Coleman_2015,Fradkin_PhysRevLett.64.1835,Polkovnikov_PhysRevB.65.064503} to the Kondo model of Eq.~\eqref{eq:mKm}. In this approach, the impurity spin is also  represented as a fermion  bilinear $\vec{S} =\bar{f}_{\sigma} \vec{s}_{\sigma\sigma^{\prime}} f_{\sigma}$. However, instead of using an imaginary chemical potential to enforce the single-occupancy constraint,  
 the latter is enforced by means of  a Lagrange multiplier, i.e. by adding
$i \lambda \sum_{\sigma} \left(\bar{f}_{\sigma} f_{\sigma} - Q\right)$ to the  action of the $f$-fermions,  Eq.~\ref{eq:spinferm}. 
Next the (spin) SU$(2)$ symmetry of the model is generalized to a larger SU$(N)$ group and, after integrating out the conduction electrons, 
it can be shown that the action in the $N\to +\infty$ limit is controlled by a saddle point, where 
$A_0 \propto J \sum_{\sigma}\langle f^{\dag}_{\sigma} d_{\sigma}\rangle/N$  is determined  self-consistently~\cite{Coleman_2015,Fradkin_PhysRevLett.64.1835}. The self-consistent equation  may not have solutions for all values of $J$, $r$, and $V$ ~\cite{Fradkin_PhysRevLett.64.1835,Polkovnikov_PhysRevB.65.064503}.   In particular, for $r=0$ (normal metal case),
there is always a solution with $A_1 = 1$ and $A_0= \rho_0 T_K$~\cite{Hewson_1993}. Thus, the saddle 
point in the $N\to +\infty$ limit maps the Kondo Hamiltonian onto resonant level model like Eq.~\eqref{eq:sK} with $U_0 = 0$, and therefore captures the physics of the singlet-doublet transitions. The  effect of the interaction term $\propto U_0$ in the Wilson ratio~\cite{Wilson_RevModPhys,Hewson_1993}, for instance, is described by including  $1/N$ corrections  to the saddle point~\cite{Read_1983}. In the present context, being able to make contact with  
the large-$N$  approach, which relies on a different small parameter, yields additional support for form of the effective  action, Eqs.~\eqref{eq:s0d}, \eqref{eq:hubbard}, and \eqref{eq:ginvc}, which is derived here using $J/D$ ($D\sim t$) as a small parameter.

Nevertheless, the ultimate validity test of the effective action from Eqs.~\eqref{eq:s0d} and \eqref{eq:hubbard} as a low-energy description of the KS rest on a renormalization-group (RG) assessment of its stability. Here we only discuss the leading order RG flow, and refer the interested reader to the more complete analysis
up to three-loops reported in Ref.~\cite{Fritz_Vojta_2004}. The starting point is the assumption that the stable RG fixed point for the KS
ground state (termed  symmetric strong-coupling, SSC fixed point in Refs.~\cite{Ingersent1_PhysRevB.54.11936,Fritz_Vojta_2004}) is described by the term $\propto A_0$ in~\eqref{eq:s0d}. This amounts to assuming a
resonant-level model with level energy tuned to the (for $r\neq 0$, singular) point at the center of the band, i.e. $\epsilon =0$.  
The terms $\propto A_1$, $V_1$ and
 $U$ are regarded as perturbations, and their leading-order RG flow follows form their (engineering)  dimension. 
 Thus, the flow equations read:
\begin{align}
\frac{dV_1}{d\ell} &= r V_1,\notag\\
\frac{dU_0}{d\ell} &= (2r-1) U_0,\notag \\
\frac{d A_1}{d \ell} &= (r-1) A_1, \label{eq:rgeq} 
\end{align}
where the dimensionless parameter $\ell$ describes the reduction of  model cutoff $\Lambda_K$ according to $\Lambda_K(\ell) = \Lambda_K e^{-\ell}$ ($\Lambda_K \sim D$ in the strong coupling regime). Notice that the additional terms that we have neglected in the derivation of the effective action involve higher number of $c$-fermion operators and/or higher powers of $\omega_n$. As it is often the case of effective field theories, this results in scaling dimensions that are higher than those of the terms proportional to $V_1,A_1$ and $U_0$, meaning that such terms only become relevant at much larger values of $r$ and therefore  have a lesser effect on the low-energy properties of the system.

The above RG equations~\eqref{eq:rgeq} imply that, whilst the SSC fixed point is perturbatively stable for $r < 0$,  for $r > 0$ the most relevant perturbation is $V_1$, i.e. PHS breaking. We will briefly discuss its effect below. However, for the moment, let us assume that  $V_1 = 0$.  In such a case, the system is at PHS and the Hubbard interaction $\propto U_0$ is the most relevant perturbation for $r > 1/2$ (for $r> 1$, $A_1$ is also relevant). 
Further analysis~\cite{Fritz_Vojta_2004} shows that, for $r > 1/2$, the SSC merges with an intermediate critical point  called symmetric critical point (SCR),  first identified by Withoff and Fradkin~\cite{Fradkin_PhysRevLett.64.1835}. For $J < J_c$, the system flows  away from the SSC/SCR into the local moment (LM) spin-doublet ground state. Moreover, for $r> 1$, the SSC/SCR is absent~\cite{Fritz_Vojta_2004} and all the flows for any $J > 0$ end up in the only stable fixed point, namely the LM. 

 In a crude attempt to understand how the ground state becomes a spin-doublet  starting from the SSC fixed point,  we may adopt the  \emph{na\"ive} approach of diagonalizing the Hubbard interaction term~\eqref{eq:hubbard} first. The latter is the most relevant term in the effective action~\eqref{eq:s0d} at PHS and $r > \tfrac{1}{2}$.  This term  favors freezing the $c$-fermion site in a singly occupied state  $c^{\dag}_{\sigma} | 0\rangle = |\sigma\rangle$ ($\sigma = \uparrow,\downarrow$), which decouples from the rest of the host at low energies, and therefore results in the spin-doublet (LM) ground state.

  Next, let us  briefly describe the case where $V_1\neq 0$ and PHS is broken. In this case, we use the same  naive approach of diagonalizing the term  $\propto V_1$ first because  it is the most relevant. This term favors  decoupling  the c-fermion site at low-energies either in the empty $|0\rangle$ (for $V_1 >0$) or doubly occupied state $|2\rangle = c^{\dag}_{\uparrow}c_{\downarrow}|0\rangle$ (for $V_1 < 0$). In either case, the ground state becomes a singlet which breaks PHS-and corresponds to the  asymmetric strong-coupling  (ASC) fixed point discussed in Refs.~\cite{Ingersent_PhysRevB.57.14254,Fritz_Vojta_2004}. However, notice that, for $r > \tfrac{1}{2}$,  the Hubbard interaction is also  a relevant perturbation, subleading to $V_1$. Therefore, we expect a competition between three possible states for the $c$-fermion site:  $\{|\sigma = \uparrow\rangle, |\sigma = \downarrow\rangle,  |0\rangle \}$, assuming $V_1 > 0$. These states correspond to a many-body doublet (LM) and a PHS-breaking (ASC)
 many-body  singlet ground state. 
 The details of the quantum criticality between these two phases cannot be obtained  using the effective theory discussed in this section, which is, strictly speaking,  valid only at PHS for $r > 0$. The correct approach is based on the infinite-$U$ Anderson model, whose RG analysis has been described in detail in Ref.~\cite{Fritz_Vojta_2004}. For full details, we refer the interested reader to  this article. Nevertheless, it is worth pointing out that the role of PHS-breaking at criticality is numerically found~\cite{Ingersent_PhysRevB.57.14254} to change at $r = r^{*} \simeq 0.375$: For $V_1\neq 0$ and $r < r^{*}$, the critical point between the LM and ASC phases is SCR~\cite{Fradkin_PhysRevLett.64.1835}, which  exhibits an emergent PHS. However, for $r >  r^{*}$, the critical point is the anti-symmetric critical point (ACR), and the spectrum of the critical system lacks PHS~\cite{Ingersent_PhysRevB.57.14254,Fritz_Vojta_2004,Zarand_PhysRevB_2025}.

Finally, we comment on the case of constant TDOS, which applies to ordinary metals with $r= 0$. For such systems,  
Nozi\'eres derived the \emph{local} Fermi liquid theory in a slightly different (but entirely equivalent) formulation from the one presented here.  For $r = 0$, $V_1$ is a marginal perturbation and  both $U_0$, $A_1$ have the same negative scaling dimension ($=-1$) and are the leading irrelevant perturbations to the KS fixed point~\cite{Wilson_RevModPhys,Nozieres_1974}.
Nozi\'eres also pointed out that, since the resonant level energy is pinned to zero energy, which for a constant TDOS is the position of the chemical potential and lacks any special meaning,  this implies that  $U_0$ and $A_1$ are in a specific numerical ratio (independent of $J$) in the regime where $J < D$~\cite{Nozieres_1974}. This results in, e.g. the Wilson ratio~\cite{Wilson_RevModPhys,Hewson_1993} taking the value of $2$, instead of unity which is expected for a non-interacting resonant level model. 

\section{Superconducting systems}\label{sec:super}

In this section, we extend the treatment presented in the previous section to superconducting systems. We treat superconductivity (or proximity to a conventional superconductor) within the mean-field approximation to the Bardeen-Cooper-Schrieffer (BCS) theory. Thus, superconducting correlations
are described by the pairing potential  $\propto \Delta$ introduced in Eq.~\eqref{eq:mKm}. When dealing with superconducting systems,  is convenient
to rewrite the quadratic parts of the Hamiltonian in the Nambu notation, e.g.:
\begin{equation}
H_T = - t \left[  \Psi^{\dag}_{d} \tau^3 \Psi_{c} + \Psi^{\dag}_{d}\tau^3 \Psi_{c} \right], \label{eq:tun}
\end{equation}
where $\tau^{\alpha=1,2,3}$ are the Pauli matrices in Nambu space and we have introduced the Nambu spinor for the $d$-fermion site:
\begin{equation}
\Psi_{d} = \left( \begin{array}{c}
d_{\uparrow}\\
d^{\dag}_{\downarrow}
\end{array}\right).
\end{equation}
In order to derive the effective low-energy field theory, we  follow  the same  steps as in Sec.~\ref{sec:normal}: i) We integrate out  the bath part of the host, which can be still done exactly because  for $\Delta \neq 0$, it is still quadratic;  ii) The the KS formed
by the magnetic impurity and the $d$-fermion site is integrated out using the cumulant expansion up to fourth order in the tunneling  amplitude $t$, see Eq.~\eqref{eq:tun}. This yields an following effective action for the $c$-fermion:
\begin{align}
S &= S_0 + S_1, \label{eq:eff}\\
S_0 &= \frac{1}{\beta}\sum_{\omega_n} \bar{\Psi}_c \left[ -\mathcal{G}_{0c}(\omega) + \Sigma_c(i \omega_n) \right] \Psi_d \notag\\
&\simeq \frac{1}{\beta}\sum_{\omega_n} \bar{\Psi}_c \left[ t^2 \tau^3 \mathcal{G}_{0d}(\omega) \tau^3 + \Sigma_c(i \omega_n) \right] \Psi_c,
\end{align}
where we have made use of the approximation~\eqref{eq:g0approx} for $\mathcal{G}_{0c}(i \omega_n)$ discussed in the Appendix~\ref{app:relation}.
The term generated
at $O(t^4)$ in the cumulant expansion has the form of the Hubbard interaction, Eq.~\eqref{eq:hubbard}, encountered earlier for
 normal Fermi systems. 
 In addition, since for superconducting systems the global gauge symmetry $U(1)$ associated with
 particle-number conservation is broken,  other types of interactions that do not conserve particle number are also permitted, but their couplings are proportional to $\Delta$ and also contain  powers of the Matsubara frequency, $\omega_n$ making them less relevant (in the RG sense) than the Hubbard interaction.
 
Furthermore, unlike the case of normal Fermi systems, 
 the Kondo-singlet self-energy $\Sigma_c(i \omega_n)$ is a matrix in Nambu space, which is calculated in Appendix~\ref{sec:zeroBW}. In addition to the term $\propto (i\omega_n \pm V)$ encountered above (the different sign applies to different components of the Nambu spinor $\Psi_c$), there is also a  local pairing potential:
 \begin{align}
 \Sigma_c(i \omega_n) &= - t^2\int d\tau e^{i\omega_n \tau} \langle  \mathcal{T}\left[ \Psi_{d}(0) \otimes \Psi^{\dag}_d(\tau) \right] \rangle_{KS} \notag\\
 &= 
 t^2 \mathcal{G}_0(i\omega_n)\notag \\
 &= 
 - \left(\frac{4t} {3J}\right)^2 \left( i \omega_n +\tau^3 V \right) - \left(\frac{4t} {3J}\right)^2 \tau^1 \Delta,
 \label{eq:sigma_sc}
 \end{align}
Hence,  the quadratic part of the action can be written as:
 \begin{equation}
 S_0= - \frac{1}{\beta}\sum_{\omega_n} \bar{\Psi}_c \mathcal{G}^{-1}_c(i \omega_n) \Psi_c,
 \end{equation}
 where
 \begin{align}
 \mathcal{G}^{-1}_c(i \omega_n) &= A_0   \left(  \frac{\omega^2_n + \Delta^2}{\Lambda_K^2}\right)^{(r-1)/2}    \left( i\omega_n   -   \tau^1 \Delta   \right) \notag\\
 &\qquad + A_1 i \omega_n  + \tau^1 B_1 \Delta  + \tau^3 V_1,
 \label{eq:gc}
 \end{align}
 where $\Lambda_K\sim D$ is a high-frequency cutoff (see below). 
 As we did for the normal Fermi system, it is interesting to note that, if we rescale the coefficients so that $A_1 \to 1$, this effective action can be  obtained by integrating out the host
 degrees of freedom of  a resonant-level model in a superconducting host. The first term describes the hopping in and out of the level. At PHS, i.e. for $V_1 = 0$ the local pairing potential $\propto B_1$  is still allowed. Indeed, for $r=0$ such a term  emerges from the Lagrange multiplier  within a large-$N$ treatment of the Kondo Hamiltonian  in superconducting host with PHS~\cite{Huang_Scipost_2024}, after setting $N=2$ and  undoing the particle-hole transformation that needs to be applied in order to generalize the spin symmetry of the Hamiltonian from SU$(2)$ to SU$(N)$ in the presence of a $s-$wave pairing potential~\cite{Huang_Scipost_2024}. 

 The most important novel feature  in Eq.~\eqref{eq:gc} brought about by superconductivity is the appearance of the local pairing term $\propto  \tau^1 B_1 \Delta $. In the strong coupling regime,  it corresponds to an additional ``path'' in the singlet-doublet fluctuations of the KS: In a virtual process, the KS is broken by scattering an electron that tunnels from the $c$-fermion site.  Thus, by virtual tunneling, the $d$-fermion site becomes doubly occupied, $|2\rangle_d = d^{\dag}_{\uparrow} d^{\dag}_{\downarrow} | 0\rangle_d$, where $|0\rangle_d$ is the empty $d$-fermion site state. The impurity state is thus a doublet. In the presence
of a paring potential $\propto \Delta$, the $d$-fermion site can coherently fluctuate between $|2\rangle_d$ and $|0\rangle_d$. Thus, at times, in order to reconstruct the KS after the virtual fluctuation, a second electron must tunnel from the $c$-fermion site. At low energies, the net effect of this process is the removal of a Cooper pair $c_{\downarrow}c_{\uparrow}$ from the $c$-fermion site. This is a local pairing term, which  was derived in  the strong coupling regime by Zazunov \emph{et al.}~\cite{Egger_PhysRevLett.121.207701} for $r = 0$. It is also generated in the weak coupling regime~\cite{Andersen_PhysRevLett.107.256802,Zazunov_PhysRevLett.103.147004}, as it can be shown using Anderson's ``Poor Man's'' scaling~\cite{Anderson_1970} for $r =0$ and $\Delta \neq 0$. In the context of the latter study, this term was related to Andreev tunneling between two superconductors. The authors of Ref.~\cite{Zazunov_PhysRevLett.103.147004} consider a single (effective) SC and called it Andreev reflection  (AR). We shall follow this naming here.  Notice that, as shown above, the AR term is, perhaps  not surprisingly, present for any value of $r$ as long as $\Delta \neq 0$. 

 Next, we can carry out an RG analysis of the AR term.  A caveat is in order, as  this analysis requires that we neglect the dependence on $\Delta$ of the term  $\propto A_0$ in Eq.~\ref{eq:gc}. This restores the scale invariance  of the term $\propto A_0$, which is a reasonable approximation as long as  the cutoff of the effective field theory remains larger than the pairing gap, i.e. $\Lambda_K \gtrsim \Delta$. Thus, assuming a small pairing potential $\Delta \ll \Lambda_K$, we can follow the RG flow  of $B_1$ all the way down to  $\Lambda_K \sim \Delta$. As in the case of $V_1, U_0$ and $A_1$ discussed in the previous section, the leading order RG flow is determined by the scaling dimension of the term $\propto B_1$, which is $r$ and yields the following flow equation:
 \begin{equation}
 \frac{d B_1}{d\ell} = r B_1.\label{eq:rgb1}
 \end{equation}
Mathematically, this equation is identical to the leading order RG flow equation for $V_1$. This is not surprising because, upon neglecting the dependence on $\Delta$ of the
term $\propto A_0$ in \eqref{eq:gc}, we can always carry out a Bogoliubov transformation with an angle $\theta = \pi/2$ and rotate $B_1$ into a term of the form of $V_1$. Notice that, in the presence of both $B_1$ and $V_1$, the Bogoliubov angle $\theta$ for the rotation will depend on both $B_1 \Delta$ and $V_1$. 

For the sake of simplicity, let us consider the system at PHS and therefore set $V_1 = 0$ in Eq.~\eqref{eq:gc}. For $r > 0$, Eq.~\eqref{eq:rgb1} means that the AR term $\propto B_1$ is the most relevant perturbation to the SSC fixed point.  Therefore, under renormalization  of $|B_1|$  grows and   favors formation of a paired state of the $c$-fermions, which we shallcall paired strong coupling,  or PSC for short. In the the gauge of the pairing potential used here~\footnote{In general, $B_1$ is a complex amplitude~\cite{Egger_PhysRevLett.121.207701}, and should appear 
as $\left(\tau^{+} B_1 + \tau^{-} B^*_1 \right)\Delta$ in \eqref{eq:gc}.} reads $|BCS\rangle = \left(|0\rangle  -  | 2\rangle\right)/\sqrt{2}$ for $B_1 > 0$ ($|\overline{BCS}\rangle = \left(|0\rangle  +   | 2\rangle\right)/\sqrt{2}$ for $B_1 < 0$), where $|2\rangle = c^{\dag}_{\uparrow}c^{\dag}_{\downarrow}|0\rangle$ and $|0\rangle$ the empty $c$-fermion site. 
However, for $r > \frac{1}{2}$ the Hubbard interaction $\propto U_0$, which favors a spin doublet is also relevant, although with a smaller (equal) scaling dimension $2r-1 < r$ for $r<1$ ($r=1$). 

Again,  we can very crudely capture the competition between the Hubbard and AR terms for $r > \tfrac{1}{2}$ by neglecting the term $\propto A_0$ in \eqref{eq:gc}, and analyzing the resulting single-site model, whose Hamiltonian (obtained from \eqref{eq:eff} after redefining $c_{\sigma} \to \sqrt{A_1} c_{\sigma}$ and $\bar{c}_{\sigma} \to \sqrt{A_1} \bar{c}_{\sigma}$ to make $A_1 = 1$ and restoring the PHS-breaking term $\propto V_1$) reads:
\begin{multline}
H_{c} = V^*_1 c^{\dag}_{\sigma}c_{\sigma} +  B^*_1 \Delta 
\left( c^{\dag}_{\uparrow} c^{\dag}_{\downarrow} + \mathrm{H.c.} \right) \\
+ U^*_0 \left(n_{\uparrow} - \tfrac{1}{2} \right)\left(n_{\downarrow} - \tfrac{1}{2} \right), \label{eq:hamc}
\end{multline}
where $V^*_1,B_1^*,U^*_0$ are  renormalized parameters at the scale where $\Lambda_K\sim \Delta$, whose values depend on $r$ and the other microscopic parameters of the model: $J, V$, etc. Eq.~\eqref{eq:hamc} is a well studied model of a magnetic impurity in a superconductor in the limit where the gap $\Delta\to +\infty$~\cite{Arovas_PhysRevB.62.6687}. This effective single-site model displays a doublet-singlet transition. The doublet state (LM) is the ground state for  $U^*_0 > 2 \sqrt{(V^*_1)^2+(B^*_1 \Delta)^2}$. The singlet is a paired state of the $c$-fermion (PSC).  At the phase boundary, there is a level crossing where the PSC and LM are degenerate in energy. Since these states have different fermion parity, the system  undergoes parity fluctuations at the transition and it close to it is therefore very prone to quasi-particle poisoning. 

 On the other hand, for $r < 1/2$, the Hubbard term $\propto U_0$ is irrelevant and therefore there is no competition of the Hubbard  and the AR terms. Thus, at strong coupling, the only stable ground state is a paired singlet (PSC) ground state. However, this does not preclude the LM (spin-doublet) ground state to be stabilized in the  weak $J$ regime, where the above effective theory does not apply. Indeed, the weak coupling analysis described in the companion  article indicates that the range of stability of the LM ground state increases with $r$. Finally, At the boundary between the LM and PSC phases, we again expect a level crossing (first order) quantum phase transition. This qualitative expectations are confirmed by the ground state phase diagrams obtained using the numerical renormalization group (NRG)~\cite{Wilson_RevModPhys,Bulla_RevModPhys.80.395}  that are described in the companion article~\cite{unpub}.

\section{NRG Results}\label{sec:nrg}

\subsection{Phase Diagram}\label{subsec:phasediag}

In the short companion paper (Ref.~\cite{unpub}), we have focused on 
the phase diagram  at or near particle-hole symmetry (PHS), i.e.  mainly for small values of the scattering potential, $V$,  and the chemical potential, $\mu$. Here we shall focus on the NRG results for arbitrarily  values of $\mu$ and $V$.  The results discussed in this section and next section have been obtained for a $s$-wave pairing potential $\Delta=10^{-3}D$

\begin{figure}
    \centering
    \includegraphics[width=\linewidth]{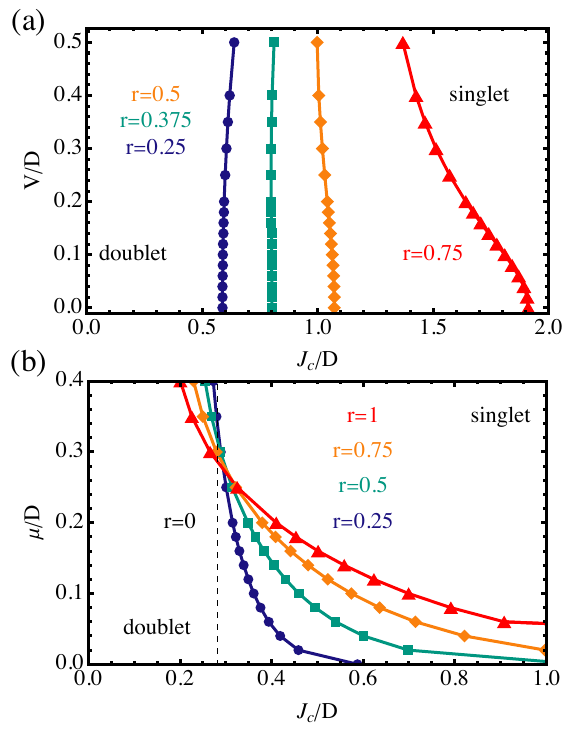}
    \caption{Doublet-singlet phase boundary (i.e. critical Kondo coupling $J_c$) as a function of (a) the strength of the    potential scattering, $V$,  and (b) the chemical potential, $\mu$,  for a $s$-wave pairing potential strength of $\Delta = 10^{-3}\:D$; $V$ ($\mu$) locally (globally)  breaks  particle-hole symmetry of the impurity model. At small values of the pseudogap exponent, $r$, the effect of $V$ is less pronounced than for  $r \gtrsim 0.5$.  For $\mu\lesssim 0.2\: D$,  the doublet-singlet phase boundaries shown on panel (b) for $r=0.25,0.5,0.75,1$ (color lines)  and  $r=0$ (dashed line) exhibit the stronger effect of the crossover from pseudogap-controlled physics to a constant tunneling density density of states. The latter results in an enhancement of $J_c$, especially for larger $r$. In both plots, $D$ denotes the system bandwidth.}
    \label{fig:phasediagV}
\end{figure}

Let us begin by considering the effects of the scattering potential $V$. As we have already pointed out~\cite{unpub}, the latter typically has a 'milder' effect on the doublet-singlet phase boundary than the chemical potential, which is also discussed below. Fig.~\ref{fig:phasediagV} (a) shows the phase boundary over a wide range of values of $V$.  The overall curvature of the boundary  changes for $r \gtrsim 0.5$. This figure also shows that  for $r\to 1$ a large $V$  substantially suppresses the critical value of the Kondo
coupling, $J_c$. On the other  hand, at small $r$, a small $V$ does not have much effect whilst a large $V$ seems to slightly favor the doublet phase. Despite the  existence of the small $s$-wave pairing potential $\Delta$ that gaps the spectrum of the host, we believe the overall behavior shown in Fig.~\ref{fig:phasediagV} is qualitatively consistent with the relevance or irrelevance (in the RG sense) of $V$ that is observed in the gapless pseudogap
Kondo system (i.e. at $\Delta = 0$).  In the gapless case and for  $r < r^*\simeq 0.375$, $V$ is irrelevant and the critical point (SCR) separating the doublet (LM) from the 
singlet (SSC) phases exhibits an emergent PHS ($V$ is relevant in the singlet phase, ASC, however)~\cite{Ingersent_PhysRevB.57.14254,Fritz_Vojta_2004,Zarand_PhysRevB_2025}.
On the other hand, for $r > r^*$, the scattering potential  $V$ is  a relevant perturbation relevant at the quantum critical point (ACR)   separating the doublet  (LM) phase from the PHS-breaking singlet (ASC) phase~\cite{Fritz_Vojta_2004,Zarand_PhysRevB_2025}.

In Fig.~\ref{fig:phasediagV}(b), we show the effect of a finite chemical potential $\mu$ on the phase boundary for different values of $r$. At $r=0$ the TDOS is constant and, the phase boundary is not sensitive to the value of $\mu$ (see dashed line on Fig.~\ref{fig:phasediagV}(b)). However, the phase boundary exhibits different behavior depending on whether $\mu < \mu^*$ or $\mu > \mu^*$ ($\mu^*\simeq0.28D$  for  $\Delta = 10^{-3} D$). In the former case (i.e. for $\mu<\mu^*$)  superconducting pseudogap systems  with $r > 0$  favor more a spin-singlet ground state than the system with constant TDOS. On the other hand, the opposite behavior is observed for $\mu>\mu^*$, that  is, the doublet is  more favored for  $\mu < \mu^*$. The effect on the phase boundary is more pronounced for $\mu < \mu^*$. Indeed, we expect the effects of the crossover from pseudogap $r > 0$  to constant (i.e. $r=0$) TDOS physics to be stronger for smaller values of $\mu$. In addition, we notice the re-entrant behavior of the singlet phase at large $\mu$ shown Fig.~\ref{fig:phasediagV}(b), which implies that the singlet-doublet level-crossing transition can be induced by tuning the chemical potential in both regimes of $\mu$.  In the following section,  we shall discuss the NRG results for the $T$-matrix spectral function and  also touch upon this chemical-potential induced transition  (cf. Fig.~\ref{fig:sp_r1}(b)).

\subsection{Spectral Function}\label{subsec:specfun}

\begin{figure}
  \centering
  \includegraphics[width=\linewidth]{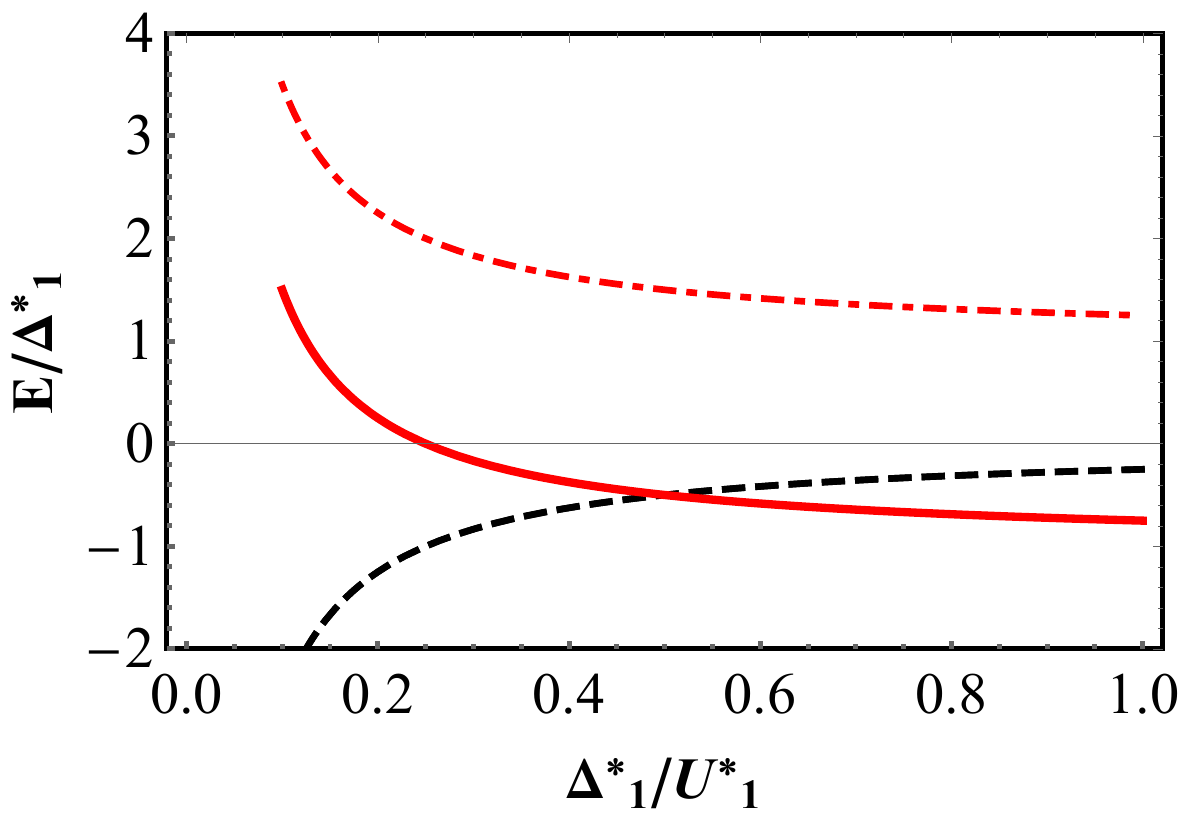}
    \caption{Spectrum of the single-site model, Eq.~\ref{eq:hamc}), at particle-hole symmetry (i.e. for $V^{*}_1 = 0$). Energies are measured in units of $\Delta^*_1 = B^*_1\Delta$, i.e. the renormalized Andreev reflection or local pairing potential; $U^*_0$ is the renormalized Hubbard interaction, arising from the singlet-triplet transitions (or Kondo-singlet polarization) in the strong coupling limit. The two low-lying states (lines) correspond to the doublet (continuous line) and singlet (dashed line). The doublet becomes the ground state for small $\Delta^*_1/U^*_0$, whilst the singlet is the ground state at large $\Delta^*_1/U^*_0$.  The upper line corresponds to the two-quasiparticle excitation. Notice that its energy is shifted down as $U^*_0$ increases. The (renormalized) Hubbard interaction, $U^*_0$, is associated with the polarization energy of the Kondo singlet and decreases with increasing $J$. }
   \label{fig:hamc}
\end{figure}

In this section, we discuss the behavior of the $T$-matrix spectral function
for the spin-$\tfrac{1}{2}$ impurity, which is accessible experimentally in tunneling
experiments. The latter is defined as follows~\cite{PhysRevLett.85.1504_Kondo}:
\begin{align}
A(\omega)  &= -\frac{1}{\pi} \mathrm{Im} \: \sum_{\sigma} T_{\sigma}(\omega),\\
T_{\sigma}(\omega) &= -i \int^{+\infty}_{0} dt \,  e^{i\omega t} \, \langle \Phi_0 \left\{ O_{\sigma}(t), O^{\dag}_{\sigma}(0) \right\} | \Phi_0\rangle,\notag\\
O_{\sigma} &= \left[d_{\sigma},  H_K  \right],\notag
\end{align}
where $H_K = J \vec{S}\cdot \vec{s}_0$ is the Kondo-exchange part of the impurity Hamiltonian.  Since, by definition, $O_{\sigma}$ is an  operator with the same  charge, spin, and fermion-parity quantum numbers as the electron $d_{\sigma}$ operator, the $T$-matrix spectral function allows  to access the spectrum of single-particle excitations. 

\begin{figure}
    \centering
\includegraphics[width=\linewidth]{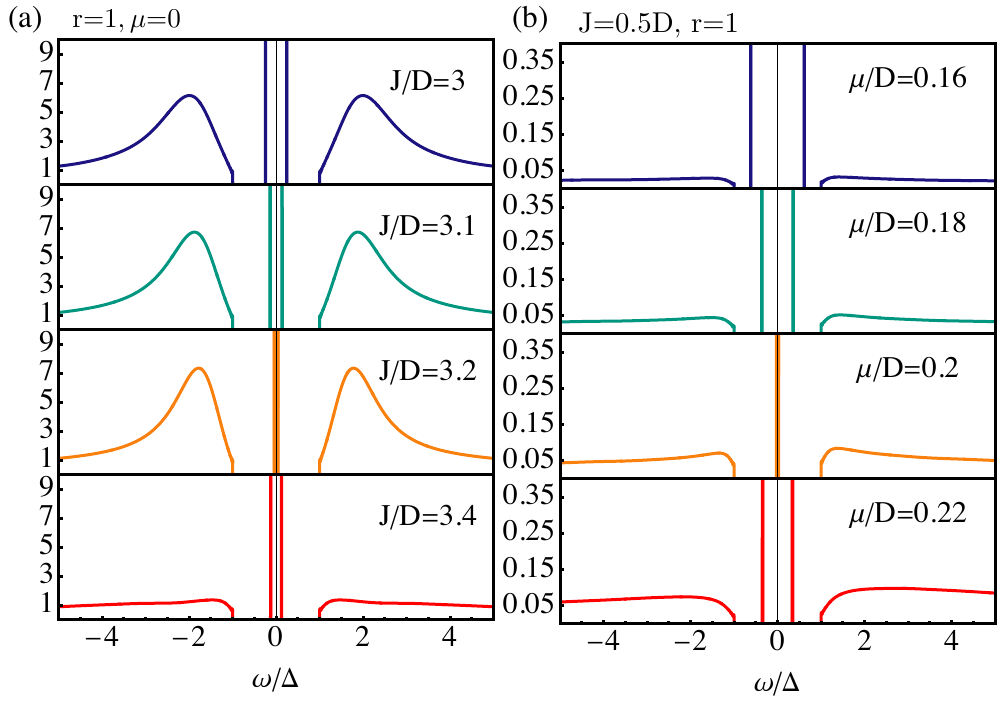}
    \caption{ Spectral function near the YSR transitions for $r=1$ driven by (a) Kondo coupling at particle-hole symmetry (i.e. for $V = \mu = 0$) and (b) chemical potential. Panel (a) also shows, besides the Yu-Shiba-Rusinov (YSR) peaks inside the gap,  a resonance-like feature for $\omega/\Delta \gtrsim 2$.  See main text for an explanation of this feature.} 
    \label{fig:sp_r1}
\end{figure}

In order to gain a deeper understanding of the single-particle spectrum  of the quantum  impurity in a SCPS host, we first focus on systems exactly at particle-hole symmetry (PHS). Although the latter may not exactly realized experimentally,  the absence of the scattering potential $V$ means that we need to keep track of fewer model parameters, which hopefully offers some simplification when trying to understand  the NRG results.  Thus 
Fig.~\ref{fig:sp_r1}(a) shows the evolution of the $T$-matrix spectral function, $A(\omega)$ (in units of the inverse bandwidth $D$), near the doublet-singlet  transition for $J/D= 3.0 \to J/D\simeq 3.4$ and $r = 1$, $V = \mu =  0$, and $\Delta/D = 10^{-3}$. Besides the expected Yu-Shiba-Rusinov (YSR) peaks within the superconducting gap, there are two prominent resonance-like features  appearing symmetrically  at $|\omega|\gtrsim  2\Delta$ for $J < J_c\simeq 3.2D$, which disappear abruptly across the transition, that is, when the ground state becomes a spin singlet (see results for $J=3.4D$ shown in the bottom panel of the the same Fig.~\ref{fig:sp_r1}(a)). Notice that,
as $J\to J_c$, the maximum of the feature appears to slightly shift down in energy and approach $\omega = 2\Delta$. Since the doublet-singlet transition in the $r = 1$ system at PHS takes place at a relatively large Kondo coupling ($J \simeq 3.2 D$),  we attempt an explanation  based on the single-site model introduced in Sec.~\ref{sec:super}, Eq.~\eqref{eq:hamc}, which we have derived from the effective field theory applying at strong coupling. Examining the spectrum of~\eqref{eq:hamc}, we see that, besides the spin-doublet ($|\sigma=\uparrow,\downarrow\rangle$) and the lowest energy
singlet ($|BCS\rangle = (|0\rangle-|2\rangle)/\sqrt{2}$ undergoing a crossing at the transition (see Fig.~\ref{fig:hamc}), there is also a higher energy spin singlet ($|\overline{BCS}\rangle = (|0\rangle+|2\rangle)/\sqrt{2}$),  which corresponds to an excited state with two Bogoliubov quasi-particles.  Notice that, 
in an extended superconductor,  such a two-quasiparticle state  lies in a continuum of other (spin-singlet) excitations with an energy threshold of $2\Delta$ and will  broaden into  a resonance-like feature rather than appearing as Dirac-delta peak (unlike the subpgap YSR states that are protected from broadening by the gap $\Delta$). 
Indeed,   examination of the spin quantum numbers of the states contributing the largest spectral weights to the resonance-like feature  seen in Fig.~\ref{fig:sp_r1}(a) reveals they are also singlets.

\begin{figure}
    \centering
\includegraphics[width=\linewidth]{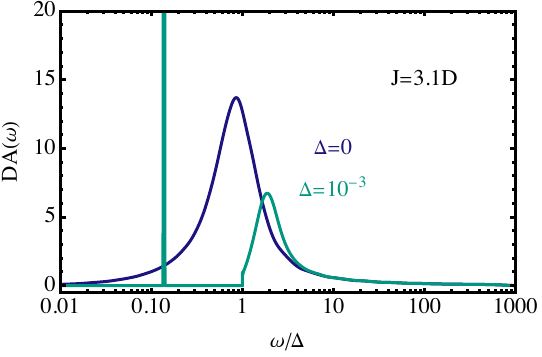}
    \caption{ Comparison  of the $T$-matrix spectrac function for the gapless ($\Delta=0$) and gapped ($\Delta=10^{-3}D$) systems for $J \to J^{-}_c$,  near the doublet-singlet phase transition at $J = J_c$. In the gapless case,   a prominent resonance-like feature appears at $\omega \simeq \Delta$. For $\Delta\neq 0$, a resonance-like feature appears at $\omega\simeq 2\Delta$ along with a Dirac-delta function peak (the YSR state) for $\omega \simeq 0$. } 
    \label{fig:gapvsgapless}
\end{figure}

 In  the model~\eqref{eq:hamc} at PHS (i.e. for $V^*_1 =0$),
the excitation energy (relative to the doublet ground state) of the two-quasiparticle
state is $\epsilon_2 = B^*_1 \Delta + U^*_0/4 - (-U^*_0/4) =  B^*_1 \Delta + U^*_0/2$ at $J \to J^{-}_c$. As explained above,  in an extended superconductor $\epsilon_2 \gtrsim 2\Delta$ for $J < J_c$. In addition, from Eq.~\eqref{eq:hamc} we can 
extract the energy gap between the (ground-state) doublet and the lowest singlet:
$\epsilon_0 =  -B^*_1 \Delta + U^*_0/4 - (-U^*_0/4) = -B^*_1 \Delta + U^*_0/2$. 
For $J \to J^{-}_c$, $\epsilon_0 \to 0$. Hence,  $B^*_1\Delta \simeq U^*_0/2$ and $\epsilon_2 = U^*_0 \simeq 2\Delta $ at the transition. However, in the gapless system for which the doublet is always the ground state (meaning that the energy gap $\epsilon_0$ never closes) and Andreev reflection is absent (i.e. $B^*_1\Delta = 0$), these equations imply that  the doublet-singlet gap is $\epsilon_0 = U^*_0/2 = \Delta$ for $J\to J^{-}_c$. Indeed,  Fig.~\ref{fig:gapvsgapless} shows a  comparison of the spectral function $A(\omega)$ for $\Delta\neq 0$ and $\Delta =0$, and $J=3.1\:D$, i.e. slightly below the transition, and shows that two resonant-like features appearing  at $\omega\gtrsim 2\Delta$ in the gaped case  
and at $\omega\simeq \Delta$ in the gapless case. Let us recall that, for $\Delta = 0$ at PHS, $U^*_0$ is the only energy scale that remains in Eq.~\eqref{eq:hamc} and it corresponds to the (twice) doublet-singlet gap.  However, for $\Delta \neq 0$  the two singlets $|0\rangle,|2\rangle$ are split  by the AR term $\propto B^*_1,\Delta$ into a lowest-energy 
singlet, $|BCS\rangle$, and the two-quasiparticle state, $|\overline{BCS}\rangle$. Therefore, doublet-doublet transition happens for the value of $J$ at which the doublet-singlet gap of the gapless system equals the $s-$wave pairing gap $\Delta$. This is sensible because the
singlet ground state will be favored when the pairing energy to be gained from the local pairing (Andreev reflection $\sim \Delta$) term overcomes this gap. 

 In addition, the model~\eqref{eq:hamc} is able to explain other numerical observations. For instance, the disappearance of  the resonance-like feature once the system crosses into the spin-singlet phase is correlated with the change in fermion parity of the ground state. Recall that the doublet and singlet ground states have opposite fermion parity and the two-quasiparticle excitation has the same parity as the   singlet. Since the operators $O_{\sigma}$ anti-commute with the fermion parity operator, single-particle excitations can only connect  two  many-body states of opposite fermion parity. This means that the two-quasiparticle state is only accessible when the ground state of the system is one of (energy degenerate) spin-doublet states. Once the system  crosses over to the spin-singlet phase, the tunneling of a single particle or hole can no longer connect the ground state and the two-quasi-particle state, which  explains why the resonance-like feature  is not seen for $J > J_c$. Furthermore, if we take into account that the physical origin of parameter $U^*_0$ is related to the polarization energy of the Kondo singlet and the latter generally  decreases with increasing  $J$, we expect the excitation energy of two-quasiparticle state,  $\epsilon_2 = B^*_1\Delta+U^*_0/2$, 
to shift down as the system approaches the transition from  the doublet phase. Indeed, we observe a (small) downward shift in the position of the maximum of the resonance-like feature in $A(\omega)$ (cf. Fig.~\ref{fig:sp_r1}(a)) as the Kondo coupling increases from $J=3.0D$ to $J=3.2D$.

 Finally, on Fig.~\ref{fig:sp_r1} panel (b) we show the results of the NRG calculation of the spectral function $A(\omega)$ for $r = 1$,  $J = 0.5 D$, and $V = 0$ and  different values of the chemical potential, $\mu$. Since $\mu$ globally breaks PHS, $A(\omega)$ is no longer symmetric 
about $\omega = 0$ (as it is the case of the spectral functions at PHS shown in panel (a)). The results shown in this panel focus on illustrating the behavior of the spectral function across the chemical-potential induced transition from the  doublet (small $\mu$) to the doublet (large $\mu$), which happens due to the re-entrant behavior of the spin-singlet phase at large $\mu$ (see Fig.~\ref{fig:phasediagV}, panel b). Apart from the imortant difference that the zero-energy crossing of the YSR is induced by $\mu$, the transition shows the same characteristics as for the system with constant TDOS (i.e. $r = 0$) as a function of $J$~\cite{Satori1992,Bulla_RevModPhys.80.395}.


\section{Conclusions}

We have studied the phase diagram and single-particle excitations of a spin-$\tfrac{1}{2}$ in a superconducting host
with $s$-wave pairing potential, $\Delta$, which also exhibits a tunneling density of states (TDOS), $\rho(\epsilon) \sim |\epsilon|^r$, for $|\epsilon|\gg \Delta$. For $\Delta = 0$ this system realizes the well-studied pseudogap Kondo model~\cite{Fradkin_PhysRevLett.64.1835,Chen_1995,Ingersent_PhysRevB.57.14254,Ingersent_PhysRevB.57.14254,Fritz_Vojta_2004,Zarand_PhysRevB_2025}. We have shown that, although  this system shares many features with its gapless version such as
the importance of presence or absence of particle-hole symmetry and different behavior for $r \lesssim \tfrac{1}{2}$
and $r \gtrsim \tfrac{1}{2}$, it also exhibits distinct features. The most important of the latter is the existence of a local pairing or ``Andreev reflection'' potential which is responsible for stabilizing a spin-singlet ground state in the regime of strong Kondo coupling for any $r$. The transition exists even when in the particle-hole symmetric case, for which the ground state of the gapless system always stays a spin doublet for $r > \frac{1}{2}$. This distinct features of the superconducting system have been explored by a combination of analytical  and numerical-renormalization group (NRG) calculations. Regarding the analytical calculations, we have presented a detailed derivation of the effective field in the strong strong coupling regime,  and discussed  the (lowest order) renormalization-group  (RG) flow of the various perturbations to it. Numerically, we have focused on the global phase diagram in the presence of particle-hole symmetry breaking perturbations such as a scattering potential or the chemical potential. We have also presented results for the $T$-matrix
spectral function, which describes the single-particle spectrum of the system. Furthermore, using an effective single-site model that we have derived from the effective field theory, we have also explained some resonant-like features observed in the spectral function in terms of a two-quasiparticle excitation of the effective single-site model. 

It is interesting to point out that, although for $r > \tfrac{1}{2}$, the effective theory discussed in Secs.~\ref{sec:normal} and \ref{sec:super} contains one or more relevant
perturbations it still retains  predictive power.  This is because although   the scattering potential $V_1$  or the Andreev reflection $B_1$ (both relevant for $r > 0$) and the Hubbard interaction $U_1$  (relevant for $r > \tfrac{1}{2}$),
are relevant perturbations in the RG sense, we can still understand the nature of the ground state by diagonalizing a model of the $c$-fermion site that contains such relevant terms only, neglecting the rest.  This approach is certainly not sufficient to understand the quantum criticality of the gapless system, but since in the present case the spectrum develops a  finite gap due to the presence of a finite $\Delta$, we believe this approach is sufficient to understand the physics
at a qualitative level. In particular, for $r > \tfrac{1}{2}$ it appears to capture well the competition between
the local pairing (Andreev reflection) term and the Hubbard term arising from the polarization of the Kondo singlet. 

Together with the results reported the companion paper, the results in this article map the regions in
parameter space where local moment (LM) ground state exhibits an enhanced stability. This can have important 
consequences for the design of superconducting spin qubits in  pseudogap superconducting  systems (SCPS). Our results
are pertinent to a single artificial impurity (SC qubit), whilst most interesting physical realizations, such minimal Majorana chains, require an array of many such qubits (impurities). Thus, an interesting and necessary direction is to study how  such quantum impurity systems couple when realized as an array of quantum dots 
coupled to the same SCPS host(s). As we have already briefly discussed in the companion paper, 
by maintaining the stability of the LM ground state at large $J$, it may be possible to engineer
strong qubit-qubit interactions, which  are necessary for qubit
logical operators and applications in quantum simulation.

\acknowledgments 

MAC thanks S. Bergeret and G. Zarand for useful discussions. CHH and MAC have been been supported by the Spanish MCIN/AEI/10.13039/501100011033 through Grant No. PID2023-148225NB-C32 (SUNRISE).

\appendix

\section{Local Matsubara Green's Function}

In this Appendix, we recall some important results for the non-interacting local single-particle Green's function for a power-law tunneling density of states (TDOS):
\begin{equation}
\rho(\epsilon) =  \rho_0 \left| \frac{\varepsilon}{D} \right|^r
\end{equation}
where $D$ plays the role of the bandwidth and $\rho_0 = (r+1)/2D$ is a constant with dimensions of inverse energy. In this work and it what follows, we assume that the TDOS is a property of the host, i.e. the system
in the absence of quantum impurity. 
In the Wilson chain representation, in the thermodynamic 
limit (i.e. for $L\to +\infty$), the above expression corresponds to  (small $\omega$ imaginary part of the  Green's function at the $m=0$ site, i.e.
\begin{equation}
\rho(\epsilon) = -\frac{1}{\pi} \mathrm{Im}\: G_{0d}(i\omega_n \to \epsilon + i 0^{+}).
\end{equation}
In order to obtain the complete form of the Matsubara Green's function $G_{0d}(i \omega_n)$
in the low-frequency limit, we reply upon the spectral representation:
\begin{align}
\label{eq:pseudogapg0}
G_{d0}(i \omega_n) &= \int d\epsilon\: \frac{\rho(\epsilon)}{i\omega_n - \epsilon}   \simeq
\frac{\rho_0}{D^r} \int^{+D}_{-D} d\nu \frac{|\epsilon|^r}{i\omega_n - \epsilon} \notag \\
&\simeq 
\left\{ 
\begin{array}{cc}
-\frac{i  \pi \rho_0  \mathrm{sgn}( \omega_n) }{\cos(\pi r/2)} \left|\frac{\omega_n}{D}\right|^r
 \qquad  & |r| < 1\\
-2i\rho_0 \left(\frac{\omega_n}{D} \right) \log\left|\frac{D}{\omega_n} \right| & r = 1
\end{array}
\right.
\end{align}
for $|\omega_n|\ll D$.

\section{Continuum Model}\label{app:cmodel}

Let us consider the following continuum model for the host, where the $c, d$-fermion sites and the bath are all described by the action  $S_c$:
\begin{align}
S &= S_c + S_K,\\ 
S_c &= \frac{1}{\beta}\sum_{k,\omega_n}  \bar{\Psi}_{k} \left[ -i \omega_n   + \xi_{k}\tau^3 + \Delta \tau^1 \right]  \Psi_{k} ,\\
S_{I}&=  \frac{\pi}{k_0 L} \sum_{k,k^{\prime}} \int d\tau \, \bar{c}_{k\sigma} \left[ \left( \vec{s}_{\sigma\sigma^{\prime}} \cdot \vec{S} \right) + V \delta_{\sigma,\sigma^{\prime}} \right] c_{k^{\prime}\sigma^{\prime}}, 
\end{align}
where $\vec{S}$ is the spin-$\tfrac{1}{2}$ operator describing the quantum impurity,  $J$ is the Kondo exchange and $V$ the scattering potential (the normalization of this term is explained below),  and  $\xi_k = \epsilon_k - \mu$ is the single-particle energy (in the normal state) referred to the chemical potential $\mu$, and $\Delta$ is the $s$-wave pairing potential (in the gauge where $\Delta$ is real), respectively. The Nambu 
spinors 
$\Psi_{k} = \left(c_{k\uparrow},\bar{c}_{-k\downarrow} \right)^T$ and $\bar{\Psi}_{k} = \left(\bar{c}_{k\uparrow}, c_{-k\downarrow} \right)$, where $c_{k\sigma}$ and $\bar{c}_{k\sigma}$  ($\sigma =\uparrow,\downarrow$) are Grassmann fields describing a set of chiral fermions with momentum $k = 2\pi (n+\tfrac{1}{2})/L$ ($n = 0,\pm 1, \pm 2$ being an integer and $L$ the channel length) and  a
power-law dispersion of the form $\epsilon_{k} = v k \left|k/k_0\right|^{\alpha-1}$, where $v$ has units of velocity and $k_0 = D/v$ is a momentum cutoff related to the host bandwidth $D$.

The exponent $\alpha$ can be related to the pseudogap exponent $r$ in  the tunneling density of states (TDOS) as follows:
\begin{align}
\rho(\epsilon) &= \frac{\pi}{k_0 L} \sum_{k} \delta(\epsilon-\epsilon_k) \\
&= \int  \frac{dk}{2 k_0}  \left|\frac{dk}{d\epsilon_k} \right| \delta(k - k_{\epsilon}) \\
&= \frac{1}{2   \alpha v k_0} \left| \frac{\epsilon}{D} \right|^{\frac{1-\alpha}{\alpha}}= \rho_0  \left| \frac{\epsilon}{D}\right|^r.
\end{align}
Hence, $\alpha =1/(r+1)$ and $\rho_0 = 1/(2 \alpha v k_0) = (r+1)/2D$. Note that the normalization is chosen such that 
\begin{equation}
\int^{+D}_{-D} d\epsilon \, \rho(\epsilon) = 1. 
\end{equation}
Using the same normalization, we introduce the  $d$-fermions, 
\begin{align}
d_{\sigma} = \sqrt{\frac{\pi}{k_0 L}} \sum_{k} c_{k\sigma},\\
\bar{d}_{\sigma}= \sqrt{\frac{\pi}{k_0 L}}  \sum_{k} \bar{c}_{k\sigma},
\end{align}
which allows to rewrite the Kondo exchange action $S_K$ in the form used in the main text (cf. Eq.~\ref{eq:mKm}): 
\begin{equation}
S_K = J \int d\tau\, \bar{d}_{\sigma} \left( \vec{s}_{\sigma\sigma^{\prime}} \cdot  \vec{S} \right)  d_{\sigma^{\prime}}
\end{equation}
In the absence of impurity, i.e. for $J = V =0$, the action is quadratic and, upon integrating all the host degrees of freedom except the $d$-fermions, the host can be described by the following action written in terms of the Nambu spinors $\Psi_d = \left(d_{\uparrow}, \bar{d}_{\downarrow}\right)^T$ and $\bar{\Psi}_d = \left( \bar{d}_{\uparrow}, d_{\downarrow}
\right)$:
\begin{align}
S_c = -\frac{1}{\beta}\sum_{\omega_n} \bar{\Psi}_d \mathcal{G}^{-1}_{0d}(i \omega_n) \Psi_d,
\end{align}
where $\mathcal{G}^{-1}_{0d}$ is the inverse of the $d$-fermion Matsubara Green's function:
\begin{align}
\mathcal{G}_{0d}(i \omega_n) &= 
-\int d\tau\,  e^{i\omega_n\tau} \langle \mathcal{T} \left[ \Psi_{d}(\tau)\otimes \Psi^{\dagger}_{d}(0) \right] \rangle_0 \notag\\
&= -\rho_0 D_r(\omega_n) \left( \frac{i\omega_n   + \tau^1 \Delta }{D}  \right),\\
D_r(\omega_n) &= 
\left\{ 
\begin{array}{cc}
\frac{\pi}{\cos(\pi r/2)}  \left(  \frac{\omega^2_n + \Delta^2}{D^2}\right)^{(r-1)/2}  & |r| < 1,\\
\ln\left|\frac{D^2}{\omega^2_n + \Delta^2} \right|   & r = 1.
\end{array}
\right. \label{eq:g00}
\end{align}
Note that, letting $\Delta \to 0$, we recover Eq.~\eqref{eq:pseudogapg0}. The same holds for $|\omega_n|/\Delta \gg 1$.

\section{Relation between $d-$ and $c-$ fermion Green's functions}\label{app:relation}

Let us  compute the form of $\mathcal{G}^{-1}_{0c}(i \omega_n)$ using the recursion relation (we assume $\mu = 0$ throughout):
\begin{align}
\mathcal{G}_{0d}(i\omega_n) &= \left[ i\omega_n  -  \Delta \tau^1  - t^2 \tau^3 \mathcal{G}_{0c}(i\omega_n) \tau^3 \right]^{-1},
\label{eq:recur2}
\end{align}
where $\mathcal{G}_{0d}(z)$ is the local Green's function of the host for $J= V = 0$, i.e. in absence of the impurity and including the $d$, $c$, and $b$-fermion sites in the tight-binding representation of Eq.~\eqref{eq:tb}. Let us stress that $\mathcal{G}_{0c}(i\omega_n)$ is not the local Green's function at $c$-fermion site of a semi-infinite chain ending at the $d$-fermion site, 
but rather the local Green's function of a  chain ending at the $c$-site (i.e. with the $d$-fermion site removed). Hence, solving \eqref{eq:recur2}  for $\mathcal{G}_{0c}(i\omega_n)$ yields
\begin{align}
\mathcal{G}_{0c}(i\omega_n) &= t^{-2}  \tau^3 \left\{ i\omega_n   -\Delta \tau^1
- \left[ \mathcal{G}_{0d}(i\omega_n) \right\}^{-1}    \right\}  \tau^3.
\end{align}
Inverting this expression we arrive at the expression:
\begin{equation}
\left[ \mathcal{G}_{0c}(i\omega_n)\right]^{-1} =  -t^2 \mathcal{F}_d(i\omega_n) \tau^3 \mathcal{G}_{0d}(i\omega_n)\tau^3 ,\label{eq:g0sc1}
\end{equation}
where
\begin{equation}
\mathcal{F}_d(i\omega_n) = \left[  1 - \tau^3 \mathcal{G}_{0d}(i\omega_n) \tau^3 \left(i\omega_n +  \tau^1 \Delta\right) \right]^{-1}.\label{eq:g0sc2}
\end{equation}
Next we recall that, at low frequencies and for a physically sensible small gap, i.e. $|\omega_n|, \Delta \ll D$, the $d$-fermion
Green's function is independent of the microscopic details of the model for the host. Thus, we insert in \eqref{eq:g0sc2} the asymptotic expression of $\mathcal{G}_{0d}(i\omega_n)$ obtained in  Appendix~\ref{app:cmodel}, Eq.~\eqref{eq:g00} using the chiral fermion representation. Thus, 
\begin{equation}
\mathcal{F}^{-1}_d(i \omega_n) = 1 - \frac{\pi  D \rho_0}{\cos(\pi r/2)} \left(\frac{\omega^2_n+\Delta^2}{D^2}  \right)^{(r+1)/2},  
\end{equation}
which is proportional to the unit matrix in Nambu space. Hence,
\begin{align}
\left[\mathcal{G}_{0c}(i \omega_n)\right]^{-1} = \frac{\pi \rho_0 t^2}{D\cos(\pi r/2)}  
\frac{\left(  \frac{\omega^2_n + \Delta^2}{D^2}\right)^{(r-1)/2} \left( i\omega_n   - \Delta \tau^1  \right) }
{1-\frac{\pi  D \rho_0}{\cos(\pi r/2)} \left(\frac{\omega^2_n+\Delta^2}{D^2}  \right)^{(r+1)/2}},
\end{align}
Thus, for $r > -1$ and $|\omega_n|, \Delta \ll D$,   the numerator dominates the behavior of Eq.~\eqref{eq:g0sc1} and therefore, we can approximate
\begin{equation}
\left[\mathcal{G}_{0c}(i \omega_n)\right]^{-1} \simeq -t^2 \tau^3 \mathcal{G}_{0d}(i \omega_n)\tau^3 .\label{eq:g0approx}
\end{equation}
This is the result used in Secs.~\ref{sec:normal} and \ref{sec:super}.Note that in the normal case, i.e. for $\Delta = 0$, $G_{0d}(i\omega_n)$ is the only independent diagonal component of $\mathcal{G}_{0d}(i\omega_n)$. 

\section{Correlations in the Kondo-singlet Ground State}\label{sec:zeroBW}

In this section, we obtain the asymptotic low-energy behavior of the correlations of the Kondo singlet consisting of the $d$- fermion  site and the spin-$\tfrac{1}{2}$  impurity. Rather than the path integral formalism used so far, this is most conveniently done in operator formalism using the spectral (Lehmann) representation of the correlation functions. 
Therefore,  let us recall the Hamiltonian of the quantum impurity plus the $d$-fermion site model:
\begin{align}
H_{I} &= J \vec{S}\cdot \left(  d^{\dag}_{\sigma}\vec{s}_{\sigma,\sigma^{\prime}}  d_{\sigma^{\prime}} \right) + \Delta \left( d^{\dag}_{\uparrow} d^{\dag}_{\downarrow} + \mathrm{H.c.}\right) + V  d^{\dag}_{\sigma} d_{\sigma} \notag\\ 
&=  J \vec{S}\cdot \vec{s}_0  + H_{sc}.  
\label{eq:onesite}
\end{align}
where $H_{sc}$ is the quadratic part of $H_I$. 
We are interested in computing the single-particle propagators of this model. For $\{A, B\} \in \{ d_{\sigma}, d^{\dag}_{\sigma^{\prime}}\}$ (fermion case), or $A,B = \left(n_{\uparrow}-\tfrac{1}{2}\right), \left(n_{\downarrow}-\tfrac{1}{2}\right)$  we want to compute the following Matsubara correlation function at zero temperature:
\begin{align}
C_{AB}(\tau) &= - \langle \mathcal{T} \left[ A(\tau) B(0) \right] \rangle_{KS}\\
&= - \langle \Phi_0| e^{H_{I} \tau} A e^{-H_{I}\tau} B | \Phi_0 \rangle \theta(\tau) \notag  \\
&\qquad \pm \langle \Phi_0 | B e^{H_{I}\tau}A e^{-H_{I}\tau} | \Phi_0\rangle \theta(-\tau) \notag \\
&= \sum_{\alpha} \left[ -\theta(\tau)  e^{(E_0-E_{\alpha})\tau} \langle\Phi_0 | A |\Phi_{\alpha}\rangle \langle\Phi_{\alpha}| B | \Phi_0\rangle\right.  \notag \\
&\quad \left.  \pm  \theta(-\tau) e^{(E_{\alpha}-E_{0})\tau} \langle\Phi_0|B|\Phi_{\alpha}\rangle \langle \Phi_{\alpha}| A|\Phi_0\rangle  \right].
\end{align}
Here $|\Phi_0\rangle$ is the ground state in the strong coupling regime where $J \gg \Delta,|V|$ and $|\Phi_{\alpha}\rangle$
are the eigenstates of $H$. In the above expressions, the $+$ sign in the second term applies to fermion operators and the $-$ sign applies to boson operators.  We are interested in the Fourier transform of $C_{AB}(\tau)$:
\begin{align}
C_{AB}(i\omega_n) &= \int d\tau  e^{i\omega_n \tau} C_{AB}(\tau)\\ 
&= \sum_{\alpha}  \left[ - A_{0\alpha} B_{\alpha 0} \int^{+\infty}_0 d\tau \, e^{\left( i \omega_n -  \omega_{\alpha 0} \right) \tau}  \right. \notag\\
&\qquad\left. \pm B_{0\alpha} A_{\alpha 0} \int^{0}_{-\infty} d\tau \, e^{\left(i\omega_n + \omega_{\alpha 0} \right)\tau} \right]\\
 &=  \sum_{\alpha} \left[ \frac{A_{0\alpha}B_{\alpha 0}}{i\omega_n - \omega_{\alpha 0}} \pm \frac{B_{0\alpha} A_{\alpha 0}}{i\omega+\omega_{\alpha 0}} \right] \\
 &= \sum_{\alpha} \left[ \frac{(A^{\dag}_{\alpha 0})^*B_{\alpha 0}}{i\omega_n -\omega_{\alpha 0}} \pm \frac{(B^{\dag}_{\alpha 0})^* A_{\alpha 0}}{i\omega_n+\omega_{\alpha 0}} \right].\label{eq:lehmann}
\end{align}
In the above expressions, we have introduced the shorthand notations $\omega_{\alpha 0} = E_{\alpha} - E_0 > 0$ and
$O_{\alpha \beta} = \langle \Phi_{\alpha} | O | \Phi_{\beta}\rangle$, etc.

In what follows, we shall illustrate the calculation of the fermionic correlation functions with $\{A,B\} \in \{d_{\sigma},d^{\dag}_{\sigma}\}$. The bosonic correlation function with $A,B = \left(n_{\uparrow}-\tfrac{1}{2}\right),\left(n_{\downarrow}-\tfrac{1}{2}\right)$ can be computed in a similar fashion, although it requires a lengthier calculation which is  prone to errors. In order to avoid such errors, we have used a well-known computer algebra program in order to obtain the exact eigenstates of $H_I$ in a symbolic form and compute the  matrix elements required for the
evaluation of Eq.\eqref{eq:lehmann}. The same script reproduces the analytical results reported below. 

In order to compute the $C_{AB}(i\omega)$ analytically we first need to diagonalize the Hamiltonian $H_{I}$ exactly. To this end,
we notice that the fermion parity $\mathcal{P} = (-1)^{\sum_{\sigma} d^{\dag}_{\sigma}d_{\sigma}}$
is a good quantum number, i.e. $\left[ H_{I}, \mathcal{P} \right] = 0$. Note that if e.g. $\{A,B\} \in \{ d_{\sigma},d_{\sigma^{\prime}}\}$, then $\{ A, \mathcal{P} \} = \{B, \mathcal{P}\} =0$, i.e. the operators $A$ or $B$ will change the parity of the state they act upon. Note that $\mathcal{P}^2 = 1$ and therefore $\mathcal{P}$ has eigenvalues $\pm 1$. In  what follows, we construct the  eigenstates in the two orthogonal subspaces where $\mathcal{P} = \pm 1$. 
\begin{table}[t]
\centering
\caption{Matrix elements for the $d$-fermion operators evaluated in the Kondo singlet ground state $|\Phi_0\rangle$.}
\label{tab:matrix_elements}
\begin{tabular}{|c|c|}
\hline
Matrix Element & Value \\
\hline\hline
\multicolumn{2}{|c|}{$d_{\uparrow}$ and $d^{\dag}_{\uparrow}$ operators} \\
\hline
$\langle \Uparrow | \otimes \langle BCS | d_{\uparrow} | \Phi_0\rangle$ & $0$ \\
$\langle \Uparrow | \otimes \langle BCS | d^{\dag}_{\uparrow} | \Phi_0\rangle$ & $-\sin(\theta/2)/\sqrt{2}$ \\
$\langle \Downarrow | \otimes \langle BCS | d_{\uparrow} | \Phi_0\rangle$ & $\cos(\theta/2)/\sqrt{2}$ \\
$\langle \Downarrow | \otimes \langle BCS | d^{\dag}_{\uparrow} | \Phi_0\rangle$ & $0$ \\
$\langle \Uparrow | \otimes \langle \overline{BCS} | d_{\uparrow} | \Phi_0\rangle$ & $0$ \\
$\langle \Uparrow | \otimes \langle \overline{BCS} | d^{\dag}_{\uparrow} | \Phi_0\rangle$ & $-\cos(\theta/2)/\sqrt{2}$ \\
$\langle \Downarrow | \otimes \langle \overline{BCS} | d_{\uparrow} | \Phi_0\rangle$ & $-\sin(\theta/2)/\sqrt{2}$ \\
$\langle \Downarrow | \otimes \langle \overline{BCS} | d^{\dag}_{\uparrow} | \Phi_0\rangle$ & $0$ \\
\hline\hline
\multicolumn{2}{|c|}{$d_{\downarrow}$ and $d^{\dag}_{\downarrow}$ operators} \\
\hline
$\langle \Uparrow | \otimes \langle BCS | d_{\downarrow} | \Phi_0\rangle$ & $-\cos(\theta/2)/\sqrt{2}$ \\
$\langle \Uparrow | \otimes \langle BCS | d^{\dag}_{\downarrow} | \Phi_0\rangle$ & $0$ \\
$\langle \Downarrow | \otimes \langle BCS | d_{\downarrow} | \Phi_0\rangle$ & $0$ \\
$\langle \Downarrow | \otimes \langle BCS | d^{\dag}_{\downarrow} | \Phi_0\rangle$ & $-\sin(\theta/2)/\sqrt{2}$ \\
$\langle \Uparrow | \otimes \langle \overline{BCS} | d_{\downarrow} | \Phi_0\rangle$ & $\sin(\theta/2)/\sqrt{2}$ \\
$\langle \Uparrow | \otimes \langle \overline{BCS} | d^{\dag}_{\downarrow} | \Phi_0\rangle$ & $0$ \\
$\langle \Downarrow | \otimes \langle \overline{BCS} | d_{\downarrow} | \Phi_0\rangle$ & $0$ \\
$\langle \Downarrow | \otimes \langle \overline{BCS} | d^{\dag}_{\downarrow}| \Phi_0\rangle$ & $-\cos(\theta/2)/\sqrt{2}$ \\
\hline
\end{tabular}
\end{table}

In addition to working in subspaces of fixed $\mathcal{P}$ eigenvalue,  it is also convenient to introduce the Bogoliubov quasi-particle operators $\gamma_{\sigma},\gamma^{\dag}_{\sigma}$, which are related to the $d$-fermion operators by means of the following canonical transformation:
\begin{align}
\Psi_d &= \left(\begin{array}{c}
d_{\uparrow} \\
d^{\dag}_{\downarrow}
\end{array} 
\right) = \left( 
\begin{array}{cc}
\cos \theta/2 & \sin \theta/2\\
-\sin\theta/2 & \cos \theta/2
\end{array}
\right) \left( \begin{array}{c}
\gamma_{\uparrow}\\
\gamma^{\dag}_{\downarrow}
\end{array}
\right) \\
&= \mathcal{U}(\theta) \left( \begin{array}{c}
\gamma_{\uparrow}\\
\gamma^{\dag}_{\downarrow}
\end{array}
\right) = U(\theta) \Gamma
\label{eq:gammabogol}
\end{align}
are the operators that diagonalize the following $2\times 2$ Bogoliubov-de Gennes Hamiltonian:
\begin{align}
\mathcal{H} &= \left( 
\begin{array}{cc}
V & \Delta \\
\Delta & -V
\end{array}
\right) = V \tau^3 + \Delta \tau^1 \\
H_{sc} &= \Psi^{\dag}_d
\mathcal{H} \Psi_d + V
\end{align}
Choosing $\tan \theta = V/\Delta$ yields
\begin{equation}
\mathcal{U}^{\dag}(\theta) \mathcal{H} \mathcal{U}(\theta) = \epsilon_{1} \tau^3.
\end{equation}
where  $\epsilon_{1} = \sqrt{V^2 + \Delta^2}$. The ground state of the quadratic part of $H_I$, i.e. $H_{sc}$, is  
the state that is annihilated by $\gamma_{0\sigma}|BCS\rangle = 0$, then 
\begin{equation}
H_{sc} | BCS \rangle = E_{0,+} |BCS\rangle,
\end{equation}
where $E_{0,+}=V  -\epsilon_1 = V - \sqrt{V^2+\Delta^2}$. Since $\{\gamma_{0\sigma},\mathcal{P} \} = \{\gamma^{\dag}_{0\sigma},\mathcal{P} \} =0$, we have that 
\begin{equation}
|\overline{BCS}\rangle = \gamma^{\dag}_{0\uparrow} \gamma^{\dag}_{0\downarrow} |BCS \rangle 
\end{equation}
is also an eigenstate of $H_{sc}$ with even fermion  parity (i.e. $\mathcal{P} =+1$) and energy $E_{1,+} = E_{0,+} + 2\epsilon_1  = V + \sqrt{V^2 + \Delta^2}$. Hence, in the even parity sector the eigenstates are the following four states:
\begin{multline}
S_{\mathcal{P}=+1} =  \left\{ |BCS\rangle \otimes |\Uparrow\rangle,  |BCS\rangle \otimes |\Downarrow\rangle,\right.\\
 \left. |\overline{BCS}\rangle \otimes |\Uparrow\rangle, |\overline{BCS}\rangle \otimes |\Downarrow\rangle\right\},
\end{multline}
with energies $E_{0,+} = V - \sqrt{V^2 + \Delta^2}$ (two-fold degenerate) and $E_{1,+} = V + \sqrt{V^2 + \Delta^2}$ (two-fold degenerate).

However, as mentioned above, in the strong coupling regime where $J\gg \Delta,|V|$, the ground state is the Kondo singlet (KS):
\begin{equation}
|\Phi_0 \rangle  = \frac{1}{\sqrt{2}} \left[ \gamma^{\dag}_{\uparrow} |BCS\rangle | \Downarrow\rangle -  \gamma^{\dag}_{\downarrow} |BCS\rangle | \Uparrow\rangle \right]
\end{equation}
This ground state is a non-degenerate spin singlet which  has an energy $E_0 = -\tfrac{3}{4} J + V$. In the odd parity sector there is also a three-fold degenerate spin triplet with $\vec{S}^2_{T} = (\vec{S} + \vec{s}_0)^2 =  2$.  However, since the fermion operators $d_{\sigma}, d^{\dag}_{\sigma}$ only change the total spin $S^z_T$ by $\pm \tfrac{1}{2}$ the triplet states  do not contribute to the single-particle propagators.  

In order to obtain the relevant matrix elements, we need the following results:
\begin{align}
d_{\uparrow} | \Phi_0 \rangle &= \left[ \cos(\theta/2) \, \gamma_{0\uparrow} + \sin(\theta/2) \gamma^{\dag}_{0\downarrow} 
\right] |\Phi_{0}\rangle \notag\\
&= \frac{1}{\sqrt{2}}  \left[ \cos(\theta/2) |BCS\rangle  - \sin(\theta/2) |\overline{BCS}\rangle \right] |\Downarrow\rangle, \\
d_{\downarrow} | \Phi_0 \rangle &= \left[ \cos(\theta/2) \, \gamma_{0\downarrow} - \sin(\theta/2) \gamma^{\dag}_{0\uparrow} 
\right] |\Phi_{0}\rangle\notag\\ 
&= \frac{1}{\sqrt{2}}  \left[ -\cos(\theta/2) |BCS\rangle  + \sin(\theta/2) |\overline{BCS}\rangle \right] |\Uparrow\rangle,  \\
d^{\dag}_{\downarrow} |\Phi_0\rangle &= \left[ -\sin(\theta/2) \gamma_{0\uparrow}+ \cos(\theta/2) \gamma^{\dag}_{0\downarrow}\right] |\Phi_0\rangle \notag\\
&= \frac{1}{\sqrt{2}} \left[-\sin(\theta/2) |BCS\rangle - \cos(\theta/2) |\overline{BCS}\rangle \right] |\Downarrow\rangle,\\
d^{\dag}_{\uparrow} |\Phi_0\rangle &= \left[ \sin(\theta/2) \gamma_{0\downarrow}+ \cos(\theta/2) \gamma^{\dag}_{0\uparrow}\right] |\Phi_0\rangle \notag\\
&= \frac{1}{\sqrt{2}} \left[-\sin(\theta/2) |BCS\rangle - \cos(\theta/2) |\overline{BCS}\rangle \right] |\Uparrow\rangle
\end{align}
Using the results in the above table combined with Eq.~\eqref{eq:lehmann}, we obtain the following expression for the single-particle Green's function where e.g. $A = d_{\uparrow}$ and $B = d^{\dag}_{\downarrow}$:
\begin{widetext}
\begin{align}
G_{0\uparrow}(i\omega_n) &= - \int d\tau e^{i\omega_n \tau} \langle \mathcal{T} \left[ d_{\uparrow}(\tau) d^{\dag}_{\uparrow}(0)\right] \rangle_{KS} = \sum_{\alpha} \left[ \frac{|(d^{\dag}_{\uparrow})_{\alpha 0}|^2}{i\omega_n - \omega_{\alpha 0}} - \frac{|(d_{\uparrow})_{\alpha 0}|^2}{i\omega_n+\omega_{\alpha 0}}\right] \notag \\
&= \frac{1}{2} \left[ \frac{\sin^2(\theta/2)}{i\omega_n - 3 J/4 - \sqrt{V^2 + \Delta^2} } + \frac{\cos^2(\theta/2)}{i\omega_n - 3 J/4 - \sqrt{V^2 + \Delta^2}} \right] \notag \\ 
&\quad +\frac{1}{2}\left[ \frac{\cos^2(\theta/2)}{i\omega_n + 3 J/4 - \sqrt{V^2 + \Delta^2}} + \frac{\sin^2(\theta/2)}{i\omega_n + 3 J/4 + \sqrt{V^2 + \Delta^2}} \right].
\end{align}
\end{widetext}
where we have used $\omega_{\alpha 0} = V - \sqrt{V^2+\Delta^2} + 3J/4 -V = 3 J/4 - \sqrt{V^2+\Delta^2}$ for $\alpha = BCS$ and $\omega_{\alpha 0} = V + \sqrt{V^2+\Delta^2} + 3J/4 - V = 3 J/4 + \sqrt{V^2+\Delta^2}$ for $\alpha = \overline{BCS}$.   In the Nambu notation, the other element in the diagonal of the propagator matrix is the correlation function 
\begin{multline}
-\int d\tau e^{i\omega_n \tau} \langle \mathcal{T} \left[d^{\dag}_{\downarrow}(\tau) d_{0\downarrow}(0) \right] \rangle_{KS} \\
= 
\int d\tau e^{i\omega_n \tau} \langle \mathcal{T} \left[d_{\downarrow}(0)c^{\dag}_{\downarrow}(\tau)   \right] \rangle_{KS}  \\
= 
\int d\tau e^{i\omega_n \tau} \langle \mathcal{T} \left[d_{\downarrow}(-\tau)d^{\dag}_{\downarrow}(0)   \right] \rangle_{KS} = - G_{0\downarrow}(-i\omega_n).
\end{multline}
In the absence of magnetic field $G_{0\uparrow}(i\omega_n) = G_{0\downarrow}(i\omega_n) = G_0(i\omega)$.

 Next, let us consider the anomalous Green's function where $A = d_{\downarrow}$ and $B = d_{\uparrow}$.
 Thus,
 \begin{widetext}
 \begin{align}
 F_{0}(i\omega_n) &= - \int d\tau \, e^{i\omega_n\tau} \langle \mathcal{T}\left[ d_{\downarrow}(\tau) d_{\uparrow}(0)\right] \rangle_{KS} = \sum_{\alpha} \left[  \frac{(d^{\dag}_{\downarrow})^*_{\alpha 0} (d_{\uparrow})_{\alpha 0}}{i\omega_n-\omega_{\alpha 0}} + \frac{(d^{\dag}_{\uparrow})^*_{\alpha 0} (d_{\downarrow})_{\alpha 0}}{i\omega_n +\omega_{\alpha 0}} \right]\\
 &= \frac{1}{2} \left[ \frac{-\cos(\theta/2)\sin(\theta/2)}{i\omega_n - 3 J/4 + \sqrt{V^2+\Delta^2}} + \frac{\cos(\theta/2)\sin(\theta/2)}{i\omega_n-3 J/4 -\sqrt{V^2+\Delta^2}} \right] \\
 & \qquad\quad + \frac{1}{2} \left[ \frac{\cos(\theta/2)\sin(\theta/2)}{i\omega_n + 3 J/4 - \sqrt{V^2 +\Delta^2}}
 + \frac{-\cos(\theta/2) \sin(\theta/2)}{i\omega_n+3 J/4 +\sqrt{V^2+\Delta^2}} \right]
 \end{align}
 \end{widetext}
 It can be seen that $F_0(i\omega) = 0$ if $\Delta = 0$, as expected. The other element of the Nambu propagator  matrix is:
 \begin{widetext}
 \begin{align}
 -\int d\tau\, e^{i\omega_n\tau} \langle \mathcal{T} \left[ d^{\dag}_{\uparrow}(\tau) d^{\dag}_{\downarrow}(0)  \right] \rangle
  &= -\int^{+\infty}_0d\tau e^{i\omega_n \tau}\, \langle d^{\dag}_{\uparrow}(\tau) d^{\dag}_{\downarrow}(0)  \rangle + 
  \int^{0}_{-\infty} d\tau \, e^{i\omega_n\tau}\, \langle d^{\dag}_{\downarrow}(0) d^{\dag}_{\uparrow}(\tau)  \rangle \\
 &= -\int^{+\infty}_0d\tau e^{i\omega_n \tau}\, \langle  d_{\downarrow}(0)d_{\uparrow}(-\tau)  \rangle^* + 
  \int^{0}_{-\infty} d\tau \, e^{i\omega_n\tau}\, \langle  d_{\uparrow}(-\tau) d_{\downarrow}(0) \rangle^* \\
  &= -\int^{+\infty}_0d\tau e^{i\omega_n \tau}\, \langle  d_{\downarrow}(\tau)d_{\uparrow}(0)  \rangle^* + 
  \int^{0}_{-\infty} d\tau \, e^{i\omega_n\tau}\, \langle  d_{\downarrow}(0) d_{\uparrow}(\tau) \rangle^*\\
  &= \left[   -\int^{+\infty}_0d\tau e^{-i\omega_n\tau}\, \langle  d_{\downarrow}(\tau)d_{\uparrow}(0)  \rangle + 
  \int^{0}_{-\infty} d\tau \, e^{-i\omega_n\tau}\, \langle  d_{\downarrow}(\tau) d_{\uparrow}(\tau) \rangle \right]^* = F^*_0(-i\omega_n).
 \end{align}
 \end{widetext}
 In this derivation we have used $\langle A(\tau) B(0) \rangle = \langle \left[ A(\tau) B(0) \right]^{\dag}\rangle^*= \langle B^{\dag}(0) \left[ A(\tau) \right]^{\dag} \rangle^*= \langle B^{\dag}(0)  A^{\dag}(-\tau) \rangle^*$. Notice that $\left[ A(\tau) \right]^{\dag} = \left[ e^{H\tau} A e^{-H\tau}\right]^{\dag} = e^{-H \tau} A^{\dag} e^{H\tau} = A^{\dag}(-\tau)$. Thus, the
Nambu matrix propagator is:
\begin{align}
\mathcal{G}_0(i\omega_m) &= - \int d\tau e^{i\omega_n \tau} \langle \mathcal{T} \left[ 
\Psi_d(0) \otimes 
\Psi^{\dag}_d(\tau) \right] \rangle \notag \\
&= \begin{pmatrix}
G_0(i\omega_n) & F_0(i\omega_n) \\
F^{*}_0(-i\omega_n) & -G_0(-i\omega_n)
\end{pmatrix}.
\end{align}
In order to simplify the expressions, is interesting to consider the form of 
$G_0(i\omega_n)$ and $F_0(i\omega_n)$ in the particle-hole symmetric case where
$V = 0$ and therefore $\cos(\theta/2) = \sin(\theta/2) = 1/\sqrt{2}$. Thus,
\begin{align}
G_0(i\omega_n) &= \frac{1}{4} \left[ \frac{1}{i\omega_n -3J/4+\Delta} +  \frac{1}{i\omega_n -3J/4-\Delta} \right] \notag\\
&+ \frac{1}{4} \left[ \frac{1}{i\omega_n + 3J/4-\Delta} +  \frac{1}{i\omega_n +3J/4-\Delta} \right] \notag\\
&=\frac{1}{2} \left[ \frac{i\omega_n}{(i\omega_n)^2 - (3 J/4 -\Delta)^2}  \right. \notag\\
&\left. \qquad\quad+  \frac{i\omega}{(i\omega_n)^2 - (3 J/4 +\Delta)^2} \right]  \notag\\
&\simeq -\frac{i\omega_n}{(3 J/4)^2} + O(\omega^3/J^3).
\end{align}
The last expression is valid in the strong coupling regime where $J\gg \Delta,\omega_n$. As for $F_0(i\omega_n)$, we have 
\begin{align}
F_0(i\omega_n) &= \frac{1}{4} \left[ \frac{-1}{i\omega_n - 3J/4 +\Delta} + \frac{1}{i\omega_n -3J/4-\Delta}\right] \notag\\
&+ \frac{1}{4} \left[ \frac{1}{i\omega + 3J/4 -\Delta} + \frac{-1}{i\omega_n +3J/4 +\Delta}\right] \\
&=\frac{1}{2}\left[ \frac{-3J/4 +\Delta}{(i\omega_n)^2 - (3J/4 -\Delta)^2}  \right. \notag\\ 
&\left.\qquad\qquad + \frac{3 J/4+\Delta}{(i\omega_n)^2 - (3 J/4+\Delta)^2} \right]\notag\\
&\simeq -\frac{\Delta}{(3J/4)^2} + O(\omega^2_n/J^2).
\end{align}
Thus, to leading order in $\omega/J$ for the particle-hole symmetric case we obtain:
\begin{equation}
\mathcal{G}_0(i\omega_n) \simeq \frac{1}{(3 J/4)^2} \left[ -i \omega_n  - \Delta \tau_1 \right].
\end{equation}
Repeating the calculation for $V\neq 0$ yields
\begin{equation}
\mathcal{G}_0(i\omega_n) \simeq \frac{1}{(3 J/4)^2} \left[ -i \omega +  \tau^3 V -  \tau^1 \Delta \right].
\end{equation}
In addition, for the bosonic correlation function we obtain:
\begin{equation}
C_{\uparrow\downarrow}(i\omega_n) \simeq \frac{1}{2J}
\end{equation}
for $|\omega_n|\ll J$.

\section{Classical Spin: YSR Approach}
\label{sec:ysr}

In the Yu-Shiba-Rusinov (YSR) approach~\cite{Yu,Shiba,Rusinov} to ingap states, the impurity spin  $\vec{S}$ is replaced by a classical vector $\vec{S} = S \hat{\vec{z}}$, where $\hat{\vec{z}}$ is the unit vector along the $z$-axis. This is expected to be a good approximation in the large $S$ limit. Thus, the Kondo exchange in the Hamiltonian becomes a local Zeeman field which in Nambu notation $V_z = \tfrac{1}{2}Js S $, where $s=\pm$, where $s^z(0)$ is the local spin density. The energy of the ingap states is computed by finding the poles of the $t$-matrix:
\begin{equation}
t_s(\epsilon) = JsS \left[ 1 - JS s \mathcal{G}^{R}_{0d}(\epsilon) \right]^{-1}
\end{equation}
where $\mathcal{G}^{R}_0(\omega) = \mathcal{G}_0(i\omega_n \to \epsilon + i 0^{+})$, with $\mathcal{G}_0(i\omega_n)$ given by Eq.~\eqref{eq:g00}. Since the term in square brackets is a $2\times 2$ matrix in Nambu space, ingap states appear when the matrix $1 - JS s \mathcal{G}^{R}_0(\epsilon)$ becomes singular, i.e. for 
\begin{equation}
\mathrm{det}\left[ 
\begin{array}{cc}
1 + g D_r(\epsilon_0) (\epsilon_0/D) & 
g D_r(\epsilon_0) (\Delta/D) \\
g D_r(\epsilon_0) (\Delta/D)  &
1 + g D_r(\epsilon_0) (\epsilon_0/D)
\end{array}
\right] = 0,
\end{equation}
where we have introduced the dimensionless coupling $g = \rho_0 J S/2$ and  set $s=+1$. Using this expression, we can find the critical value of $g \propto J$ for which $\omega_0 = 0$. In the YSR approach~\cite{Yu,Shiba,Rusinov}, this is the condition for the so-called parity-changing quantum phase transition, and for general $r$   (assuming $g_c > 0$) we have:
\begin{equation}
g_c  D_r(\epsilon_0 = 0)  \left( \frac{\Delta}{D} \right) = 1.
\end{equation}
Specializing to the case of constant (normal-state) density of states where $r=1$ this leads the well known results $\pi g_c = 1$,
i.e. $J_c  = 2/(\pi \rho_0)\sim D$, which is independent of the strength of the pairing potential $\Delta$. For $r\neq 0$, $J_c$ depends on $\Delta$. In particular, for superconducting Dirac systems where $r=1$, we arrive at:
\begin{equation}
g_c = \tfrac{1}{2}\rho_0 J_c = \frac{D}{2\Delta \ln \left|\frac{D}{\Delta} \right|},
\end{equation}
which agrees with the result derived in Ref.~\cite{Loss_PhysRevB_2014} for a magnetic impurity on a  topological insulator surface in close proximity to
 a conventional superconductor. For instance, for $\Delta/D =10^{-2}$, 
\begin{equation}
\rho_0 J_c \simeq 100/4.61 \simeq 21.71, 
\end{equation}
which is an order of magnitude larger than the critical coupling for a system with constant density of states. 

\section{Superconducting Graphene or 3D Topological Insulator Surface}\label{app:scps}

Let us compute the TDOS for superconducting graphene. In the long wave-length
limt, the $\pi$-band states can be described by the following $\vec{k}\cdot \vec{p}$ Hamiltonian after diagonalization
in the valence/conduction basis:
\begin{equation}
H_0 =  \sum_{\vec{k},\lambda} \psi^{\dag}_{\vec{k},\sigma\lambda} \left( \epsilon_{\vec{k}}\gamma^3 - \mu \right) \psi_{\vec{k},\sigma\lambda}.
\end{equation}
Here  $\epsilon_{\vec{k}} = v_F |\vec{k}|$, for $|\epsilon_{\vec{k}}| < D$, $\lambda $ is the valley index ($\lambda=-\pm 1$ for graphene and $\lambda = -\lambda = 0$ for 3D TI surfaces),  $\gamma^3 = \mathrm{diag}(1,-1)$ is the  $2\times 2$  helicity matrix of the spinor:
\begin{equation}
\psi_{\vec{k},\sigma\lambda} = \left( 
\begin{array}{c}
c_{\vec{k},\sigma\lambda}\\
v_{\vec{k},\sigma\lambda}
\end{array}
\right)
\end{equation}
Where $c_{\vec{k},\sigma}$ ($v_{\vec{k},\sigma}$ ) creates
an electron in the conduction (valence) band with  momentum $\vec{k}$ and spin $\sigma =\uparrow,\downarrow$.
The $s$-wave pairing potential induced by proximity takes the form:
\begin{equation}
H_{\Delta} = \sum_{\vec{k},\lambda} \left[ \Delta \psi^{\dag}_{\vec{k},\uparrow \lambda} \left( \psi^{\dag}_{-\vec{k},\downarrow, -\lambda} \right)^T + \mathrm{H.c.}\right]
\end{equation}
Introducing the following Nambu spinors:
\begin{equation}
\Psi_{\vec{k},\lambda} = \left( 
\begin{array}{c}
\psi_{\vec{k},\uparrow,\lambda} \\
\left( \psi^{\dag}_{-\vec{k},\downarrow,-\lambda} \right)^T
\end{array}
\right) = \left( 
\begin{array}{c}
c_{\vec{k},\uparrow,\lambda}\\
v_{\vec{k},\uparrow,\lambda}\\
c^{\dag}_{-\vec{k},\downarrow,-\lambda}\\
v^{\dag}_{-\vec{k},\downarrow,-\lambda}
\end{array}
\right)
\end{equation}
we can rewrite the total Hamiltonian as follows:
\begin{align}
H = H_0 + H_{\Delta} = \sum_{\vec{k},\lambda} \Psi^{\dag}_{\vec{k}\lambda} \mathcal{H}_{BdG}(\vec{k}) \Psi_{\vec{k}\lambda},
\end{align}
where the Bogoliubov-de Gennes Hamiltonian reads:
\begin{equation}
\mathcal{H}_{BdG}(\vec{k}) = \left( \epsilon_{\vec{k}} \gamma^3 - \mu\right) \tau^3  + \Delta \tau^1
\end{equation}

Next, let us compute the local Matsubara Green's function, i.e.
\begin{align}
\mathcal{G}_{0d}(\omega_n) &= \frac{1}{N}  \sum_{\vec{k}} \mathrm{Tr}_{BV}  \left( i\omega_n - \mathcal{H}_{BdG}(\vec{k}) \right)^{-1} \notag\\
 &= a^2 \sum_{\gamma=\pm} \int \frac{d\vec{k}}{(2\pi)^2}  \left[ \omega_n - (\gamma \epsilon_{\vec{k}} - \mu)\tau^3 - \Delta \tau^1  \right]^{-1} \notag \\
 &=- a^2 \sum_{\gamma=\pm,\lambda}   \int \frac{d\vec{k}}{(2\pi)^2}  \frac{i\omega_n + (\gamma \epsilon_{\vec{k}}-\mu)\tau^3+\Delta \tau^1}{\omega^2_n + (\gamma \epsilon_{\vec{k}} - \mu)^2 + \Delta^2}
\end{align}
In the first line of the previous equation,  $N$ is the number of unit cells of the graphene/TI host, and $\mathrm{Tr}_B\left[\ldots \right]$ stands for partial
trace over the  bands, that is, the eigenvectors of the helicity matrix $\gamma^3$, which
has eigenvalues $\pm 1$, and the valleys (i.e. over $\lambda$). Notice that we do not trace over the
Nambu indices. In addition, in the second line $a$ is the lattice parameter. 

Next, we specialize to the case where $\mu = 0$, which yields
\begin{align}
\mathcal{G}_{0d}(\omega_n) &=  -a^2 g_V g_B \int^{+D/v_F}_{0}  \frac{k dk}{2\pi}  \frac{i\omega_n + \Delta \tau^1}{\omega^2_n + (v_F k)^2 + \Delta^2}\\
&=   -a^2 \frac{g_V g_B}{2\pi v^2_F} \int^{+D}_{0} \xi d\xi  \frac{i\omega_n + \Delta \tau^1}{\omega^2_n +  \xi^2 + \Delta^2}
 \\
&\simeq- \rho_0 \left( \frac{i\omega_n + \Delta \tau^1}{D} \right)  \ln \left| \frac{D^2}{\omega^2_n + \Delta^2} \right|,
\end{align}
where $\rho_0 = g_V g_Ba /4\pi v_F = g_V g_B/D$, with $g_B = 2$ and $g_V = 2$ for graphene and $g_V = 1$ for 3D TI surfaces, and $D = v_F/a$. 
In the last line we have assumed that $|\omega_n|, \Delta \ll D$.

\section{$d+ i s$-wave Superconductors}\label{app:scps2}

 Let us  briefly review the argument provided in Ref.~\cite{Fritz_PhysRevB_2005} about the role of anomalous propagators in the  Kondo effect of magnetic impurities in 
 unconventional superconductors.   The starting point is the Anderson model describing a general type of magnetic impurity in an unconventional superconductor (treated within the BCS approximation):
 \begin{align}
 S &= -\frac{1}{N \beta} \sum_{\vec{k},\omega_n}\bar{\Psi}_{\vec{k}} 
 G^{-1}_0(\vec{k},\omega_n) \Psi_{\vec{k}}(\omega_n) \notag\\
 &+ \frac{1}{\beta}\sum_{\omega
 _n} \bar{\Psi}_f \left(-i\omega_n  + \epsilon_d \tau^3 \right) \Psi_f + S_{\mathrm{int}}(\bar{\Psi}_f,\Psi_f) \notag\\
 &+\frac{1}{\beta\sqrt{M}} \sum_{\vec{k},\omega_n}\left[ \bar{\Psi}_{\vec{k}} V_{\vec{k}} \tau^3 \Psi_f +    \bar{\Psi}_f \left(V_{\vec{k}} \tau^3 \right)^{\dagger} \Psi_{\vec{k}} \right], 
 \end{align}
where the sums over the lattice wavevector $\vec{k}$ are restricted to the 1st  Brillouin zone  and the Nanbu spinor of  Grassmann fields $\Psi^{T}_{\vec{k}}(\tau) = \left( c_{\vec{k}\uparrow}(\tau), \bar{c}_{-\vec{k},\downarrow}(\tau) \right)$ describes the degrees of the superconductor. Likewise, $\Psi_f^T(\tau) = \left(f_{\uparrow}(\tau), \bar{f}_{\downarrow}(\tau)\right)$, describes the local  degrees of freedom of the impurity. The  self-interaction of the latter is described by the standard Hubbard-$U$ term, i.e.
\begin{equation}
S_{\mathrm{int}}(\bar{\Psi}_f,\Psi_f) = 
\frac{U}{\beta}\sum_{\omega_n} \bar{f}_{\uparrow} \bar{f}_{\downarrow} f_{\downarrow} f_{\uparrow}.
\end{equation}
In the above action, we have also introduced the tunneling matrix:
\begin{equation}
V_{\vec{k}} =  \sum_{\vec{R}} 
 \left(
\begin{array}{cc}
e^{i \vec{k}\cdot \vec{R}} V_{\vec{R}} & 0 \\
0 & e^{i \vec{k}\cdot \vec{R}} V^{*}_{\vec{R}}
\end{array}
\right). 
\end{equation}
Within the BCS mean-field approximation, the superconductor degree's of freedom are not interacting and entirely described by the  Green's function:
\begin{equation}
G^{-1}_0(\vec{k},\omega_n)  = 
\left( 
\begin{array}{cc}
i\omega_n - \xi_{\vec{k}} & \Delta_{\vec{k}} \\
\Delta^*_{\vec{k}} & i\omega_n + \xi_{\vec{k}}
\end{array}
\right)
\end{equation}
where $\Delta_{\vec{k}}$ is  the BCS pairing potential, which we will take to be of the $d_{x^2-y^2} + i s$ symmetry:
\begin{equation}
\Delta_{\vec{k}} = \Delta_s  + i \Delta_d  f_d(\vec{k}),
\end{equation}
where $f_d(\vec{k}) = (\cos k_x a - \cos k_y a)$ if we  assume superconducting host to be a material with a square lattice of lattice parameter $a$. For the $s$-wave component, we can also assume a $k$-dependent $A_{1g}$ form $\Delta_s(\vec{k}) = \Delta_s f_s(\vec{k})$ where  $f_s(\vec{k}) = (\cos k_x a + \cos k_y a)$.  However, using this form will not lead to a qualitatively different result, and therefore we stick to a the simpler constant $\Delta_s$ form.

In order to derive a local description of this impurity problem, we first integrate out the degrees of freedom of the superconductor, which yields the following action:
\begin{align}
S& = \frac{1}{\beta}\sum_{\omega
 _n} \bar{\Psi}_f \left(-i\omega_m  + \epsilon_d + V^{2} \tau^3 \mathcal{G}_0(\omega_n) \tau^3 \right) \Psi_f \notag\\
 &\quad + S_{\mathrm{int}}(\bar{\Psi}_f,\Psi_f)
\end{align} 
where we have denoted:
\begin{align}
\mathcal{G}_0(\omega_n) &= \frac{1}{N }\sum_{\vec{k}}    H^{\dagger}_{\vec{k}}  G_0(\vec{k},\omega_n)    H_{\vec{k}} \\
&=\sum_{\vec{k}} 
\left(
\begin{array}{cc}
|h_{\vec{k}}|^{-2}  \left(i\omega_n - \xi_{\vec{k}}\right)  & h^{-2}_{\vec{k}} \Delta_{\vec{k}} \\
(h^{*}_{\vec{k}})^{-2} \Delta_{\vec{k}} & |h_{\vec{k}}|^{-2}  \left(i\omega_n 
+\xi_{\vec{k}}\right)^{-1}
\end{array}
\right)
 \end{align}
and introduced $V^2 = \sum_{\vec{R}}|V_{\vec{R}}|^2$ and $H_{\vec{k}} = V_{\vec{k}}/V = 
\mathrm{diag}\left\{ h_{\vec{k}}, h^*_{\vec{k}} \right\}$, 
$h_{\vec{k}}  = \sum_{\vec{k}} e^{i\vec{k}\cdot \vec{R}} V_{\vec{R}}/V$. 

To make further analytical progress, we assume the impurity coupled with a ($s$-wave) hybridization $V_{\vec{R}}$  to its nearest neighbors, i.e. $V_{\vec{R} = \pm\vec{a}_x} =  V_{\vec{R} = \pm\vec{a}_y} = V$, where $\vec{a}_x = (a, 0)$ and $\vec{a}_y = (0, a)$ , which implies that $h_{\vec{k}} = 1$.   In this case, notice that under a $\pi/2$  ($C_4$) rotation, the $d_{x^2-y^2}$ part of $\Delta_{\vec{k}} \propto  \Delta_d f_d(\vec{k})$ changes sign whilst $f_s(\vec{k})$ does not. Thus, upon summation over $\vec{k}$, the of $d$-wave part of $\Delta_{\vec{k}}$ yields a vanishing contribution to the off diagonal (anomalous)
components of $\mathcal{G}_0(\omega_n)$~\cite{Fritz_PhysRevB_2005}. The latter only receive contributions from the $s$-wave component $\propto \Delta_s$. 

 For $\Delta_s  = 0$,  the hybridization part of the action is entirely
 determined by  the diagonal (normal) elements of $\mathcal{G}_0(\omega_n)$ and follows from the
 the density of states,  which is linear in energy, i.e. $\rho(\epsilon) = - \mathrm{Im}
\left[ \mathcal{G}_{0d} (i\omega_n \to \epsilon + i 0^{+})\right]_{11}/\pi \rho(\epsilon) \propto |\epsilon|$ at low energies~\cite{Maki2001} for energies $|\omega|\ll \Delta_s$. This is a consequence of the presence of linearly-dispersing quasi-particles in the spectrum of $d$-wave superconductors~near the nodal points where $\Delta_{\vec{k}}$ vanishes~\cite{Maki2001}. On the other hand, for $\Delta_s \neq 0$ the quasi-particle spectrum is fully gapped an both diagonal and nondiagonal elements of $\mathcal{G}_0(\omega_n)$ contribute to the hybridization. Assuming a low lattice filling which implies a near circular (normal state) Fermi surface and approximating $f_d(\vec{k})
\simeq \cos 2\theta$, where $\theta$ is the polar angle, we obtain:
\begin{align}
\mathcal{G}_{0d}(\omega_n) 
&\simeq   \int^{+\pi}
_{-\pi} \frac {d\theta}{2\pi} \int^{+D}_{-D} d\xi  
\frac{-\nu_0}{\omega^2_n+\xi^2+\Delta^2_s + \Delta^2_d \cos^2(2\theta)} \notag\\
&\times
\left(
\begin{array}{cc}
i\omega_n + \xi & \Delta_s + i \Delta_d \cos 2\theta \\
\Delta_s - i \Delta_d \cos 2\theta & i\omega_n - \xi 
\end{array}
\right) \\
&\simeq  \int^{+\pi}
_{-\pi} \frac {d\theta}{2\pi}
\frac{-\pi \nu_0 \left(i\omega_n + \Delta_s \tau^1\right)}{\sqrt{\omega^2_n+\Delta^2_s + \Delta^2_d \cos^2(2\theta)}},\\
\end{align}
where $\nu_0$ is the normal state density of
states.  Finally, we shall further assume that the $s$-wave component of the pairing potential is small, i.e. $\Delta_s\ll \Delta_d$, 
we can obtain the asymptotic
behavior of $\mathcal{G}_0(i\omega_n)$ in the limit where
$\omega^2_n + \Delta^2_s\ll \Delta^2_d$ is small. In this limit
\begin{equation}
\mathcal{G}_{0d}(\omega_n) \simeq - 4\nu_0 \left(\frac{i\omega_n + \Delta_s \tau^1}{4\Delta_d} \right) 
\log\left|\frac{16\Delta^2_d}{\omega^2_n+\Delta^2_s} \right|.
\end{equation}
Notice that this becomes of the form \eqref{eq:g00} with $r=1$  if we identify $\rho_0 = 4\nu_0$ and $D = 4\Delta_d$ and $\Delta = \Delta_s$.

\section{Details of the NRG calculations}\label{app:nrg}

The Hamiltonian of $d+is$ superconductor described in the main text can be represented as a (Wilson chain) 1D lattice using the scheme introduced in Ref.~\cite{PhysRevB.72.104432_Campo} for the density of states $\rho(\epsilon-\mu)= |\epsilon-\mu|^r/E_0$ in the interval $[-D, D]$ where $E_0 =\int_{-D}^D\rho(\epsilon-\mu)\:d\epsilon$ is the renormalization factor. The discretization parameter is taken to be $\Lambda=2$. This procedure yields the Wilson chain Hamiltonian, $H  =H_\text{hyb}+H_\text{cb}$ where,
\begin{align}
H_\text{hyb}&=J\vec{S}\cdot\vec{s}(0)+V\sum_\sigma f^\dag_{0\sigma} f_{0\sigma},\\
H_\text{cb}&=\sum_{i\geq 0} \biggl\{ t_i\sum_\sigma \left[f_{i,\sigma}f^\dag_{i+1,\sigma}+ \mathrm{H.c.} \right]\notag \\
&+\epsilon_i\sum_\sigma\left[f^\dag_{i,\sigma}f_{i,\sigma}\right]+ \Delta\left[f^\dag_{i\uparrow} f^\dag_{i\downarrow} + \mathrm{H.c.}\right]\biggr\},
\end{align}
where the hopping decays exponentially as $t_N\sim \Lambda^{-N/2}$. Notice that the pairing potential term does not conserve particle number Therefore, the conserved quantum number is the U($1$) spin projection $S_z=S_{z;imp} + \frac{1}{2}\sum_{i} \left(f^\dag_{i\uparrow} f_{i\uparrow}-f^\dag_{i\downarrow} f_{i\downarrow}\right)$. The Hamiltonian can be solved with standard NRG routines. 

The computation can be made efficient for the particle-hole symmetrical case, $V=0,\mu=0$. In this limit, the Hamiltonian preserves the U($1$) charge quantum number $Q_x = \sum_i\left[f^\dag_{i\uparrow} f^\dag_{i\downarrow} + \mathrm{H.c.}\right]$~\cite{PhysRevB.90.241108_YAO,TBQ_huang_2026}. As described in the main text, we can apply  a Bogoliubov transformation~\cite{Satori1992}:
\begin{align}
b^\dag_{i,\uparrow}&=\frac{1}{\sqrt{2}} \left(f^\dag_{i,\uparrow}+f_{i,\downarrow}  \right),\\
b_{i,\downarrow}&=\frac{1}{\sqrt{2}} \left(f^\dag_{i,\uparrow}- f_{i,\downarrow}  \right).
\end{align}
followed by a particle-hole transformation,
\begin{align}
c^\dag_{2i,\uparrow}&=b^\dag_{2i,\uparrow},\\
c_{2i,\downarrow}&= b_{2i,\downarrow},\\
c^\dag_{2i-1,\uparrow}&=b_{2i-1,\downarrow},\\
c_{2i-1\downarrow}&=-b^\dag_{2i-1,\uparrow}.
\end{align}
Thus, we arrive at the following model:
\begin{align}
H &= \sum_{i\geq 0} \Big\{t_i \sum_{\sigma}\left( c_{i,\sigma} c_{i+1,\sigma}^\dag+ \mathrm{H.c.} \right) \notag\\&+ \Delta   (-1)^i Q^z_i   \Big\}+J \vec{S}\cdot\vec{s}_{0}.
\end{align}
In this new basis, the NRG is applied using the following conserved $U(1)$ quantities $(Q^z,S^z_T)$: 
\begin{align}
Q^z&=\sum_{i} Q^z_i = \sum_i \left[ n_{i,\uparrow}+n_{i,\downarrow}-1 \right] ,\\
S^z_T &=S^z_\text{imp} + \sum_i s^z(i)= S^z_\text{imp}+\frac{1}{2}\sum_i \left[n_{i,\uparrow}-n_{i,\downarrow}\right],
\end{align}
which is suitable for the case with external magnetic fields. At each iteration, we keep at least $1024$ states and discard the states above the energy scale $\omega \approx 10\omega_N=10 \Lambda^{(1-N)/2}$. In the presence of the (superconducting) gap, the NRG iteration must be truncated at iterations with energy scale $\omega_N\ll \Delta$~\cite{Hecht_2008_BCS}. Thus, we stop our NRG computation at iterations with energy scale $\sim 10^{-5}\Delta$ which is sufficient to obtain the spectral properties accurately. We set the temperature $T\ll\Delta$ so effectively that we can consider our results to be in the zero-temperature limit.  For the Kondo model, the spectral function, $W_\sigma(\epsilon)$, are defined using the T-matrices \cite{PhysRevLett.85.1504_Kondo}. The spectral weight reads,
\begin{align}
W_\sigma(\epsilon) &= -\frac{1}{\pi}\text{Im}\:  C^{R}_{\sigma}(\epsilon),\\
C^{R}_{\sigma}(\epsilon) &= \int dt \, e^{i\epsilon t} C^{R}_{\sigma}(t),\\
C^{R}_{\sigma}(t) &= -i \theta(t) \langle\{ O_{\sigma}(t), O^{\dag}_{\sigma}(0) \} \rangle 
\end{align}
 where $O_\sigma = [ f_{0\sigma}, H_K]$ with $H_K = J \vec{S}\cdot \vec{s}_0$. Note that $f_{0\sigma}$ is the operator in the original  fermion basis of the superconducting model. To carry out the computation, we obtain the spectral weights using the full density-matrix scheme described in Ref.~\cite{PhysRevLett.99.076402_FDM} and broaden the discrete data set using a hybrid kernel. The spectral function, $A_{\sigma}(\omega)$ is thus computed from the following expression:
\begin{align}
A_\sigma(\omega) &= \sum_\epsilon W_\sigma(\epsilon)\biggl\{\Theta(\epsilon)\Big[\Theta( \epsilon-\epsilon_\text{gap}^{+} ) lG(\omega,\epsilon,a)\notag \\ &+ \Theta( \epsilon_\text{gap}^{+}-\epsilon)G(\omega,\epsilon,b)\Big],\notag
\notag\\
&+\Theta(-\epsilon)\Big[\Theta( \epsilon_\text{gap}^{-} -\epsilon) lG(\omega,\epsilon,a)\notag \\ &+ \Theta(\epsilon- \epsilon_\text{gap}^{-})G(\omega,\epsilon,b)\Big]
\biggr\}.
\end{align}
where
\begin{align}
&lG(\omega,\epsilon,a)\notag\\
&=\frac{\Theta(\omega\epsilon)}{a|\omega|\sqrt{\pi}}\text{Exp}\left[ - \left( \frac{\log(|\omega|)-\log(|\epsilon|)}{a}-\frac{a}{4} \right)^2\right],\\
&G(\omega,\epsilon,b)=\frac{1}{b\sqrt{\pi}}\text{Exp}\left[{-\left(\frac{\epsilon-\omega}{b}\right)^2}\right].
\end{align}
 $\epsilon^+_\text{gap}$ and $\epsilon^-_\text{gap}$ are the positions of the BCS gap at positive and negative sides. They are determined from the data of spectral weights. Outside the gap, we use a logarithmic mesh binning $\sim 500$ points per decade with respect to the gap and a Log-Gaussian kernel with a broadening parameter, $a=0.4$. Inside the gap, we accumulate all the spectral weights and broaden the weights using a Gaussian kernel with width $b=\Delta/1000$. To eliminate the oscillatory artifacts in the continuum due to discretization, the spectral functions are z-averaged~\cite{PhysRevB.41.9403_Z} using $4$ z-points spanning the interval $[1/4,1]$.

%
\bibliography{references,NRG}

%
%
%

\end{document}